\documentclass{article}

\usepackage{arxiv}

\usepackage[utf8]{inputenc} 
\usepackage[T1]{fontenc}    
\usepackage{hyperref}       
\usepackage{url}            
\usepackage{booktabs}       
\usepackage{amsfonts}       
\usepackage{nicefrac}       
\usepackage{microtype}      
\usepackage{lipsum}
\usepackage{graphicx}
\usepackage{natbib}
\usepackage{amsmath}
\graphicspath{ {./images/} }

\usepackage[english]{babel}
\newtheorem{theorem}{Theorem}
\newtheorem{lemma}{Lemma}

\title{Robustness intervals for competing risks analysis with causes of failure missing not at random}

\author{
 Giorgos Bakoyannis \\
  Department of Biostatistics and Health Data Science \\
  Indiana University Indianapolis\\
  Indianapolis, Indiana, U.S.A. \\
  \texttt{gbakogia@iu.edu} \\
   \And
 Aristofanis Rontogiannis \\
  Department of Biostatistics \\
  Brown University \\
  Providence, Rhode Island, U.S.A. \\
  \texttt{aristofanis$\_$rontogiannis@brown.edu} \\
  \And
 Ying Zhang \\
  Department of Biostatistics \\
  University of Nebraska Medical Center\\
  Omaha, Nebraska, U.S.A. \\
  \texttt{ying.zhang@unmc.edu} \\
  \And
 Wanzhu Tu \\
  Department of Biostatistics and Health Data Science \\
  Indiana University Indianapolis \\
  Indianapolis, Indiana, U.S.A. \\
  \texttt{wtu1@iu.edu} \\
  \And
 Ann Mwangi \\
  Department of Mathematics, Physics, and Computing \\
  Moi University \\
  Eldoret, Kenya \\
  \texttt{annwsum@gmail.com} \\
  \And
 Constantin T. Yiannoutsos \\
  Department of Epidemiology and Biostatistics \\
  City University of New York \\
  New York City, New York, U.S.A. \\
  \texttt{Constantin.Yiannoutsos@sph.cuny.edu} \\
}

\begin{document}
\maketitle
\begin{abstract}
Analysis of competing risks data is often complicated by the incomplete or selectively missing information on the cause of failure. Standard approaches typically assume that the cause of failure is missing at random (MAR), an assumption that is generally untestable and frequently implausible in observational studies. We propose a novel sensitivity analysis framework for the proportional cause-specific hazards model that accommodates missing-not-at-random (MNAR) scenarios. A sensitivity parameter is used to quantify the association between missingness and the unobserved cause of failure. Regression coefficients are estimated as functions of this parameter, and a simultaneous confidence band is constructed via a wild bootstrap procedure. This allows identification of a range of MNAR scenarios for which effects remain statistically significant; we refer to this range as a robustness interval. The validity of the proposed approach is justified both theoretically, via empirical process theory, and empirically, through simulation studies. We apply the method to the analysis of data from an HIV cohort study in sub-Saharan Africa, where a substantial proportion of causes of failure are missing and the MAR assumption is implausible. The analysis shows that key findings regarding risk factors for care interruption and mortality are robust across a broad spectrum of MNAR scenarios, underscoring the method’s utility in situations with MNAR causes of failure.
\end{abstract}


\section{Introduction \label{intro}}


Competing risks data are common in many research settings, including observational cohort studies and clinical trials. Analysis is often complicated by incomplete information on the event type, or cause of failure, for a subset of study participants. Causes of missingness include data loss, unrecorded event classifications, and reporting delays. Na\"ive analytical approaches, such as complete case analysis, or treating missing causes as an additional failure category, can lead to substantially biased estimates and questionable inferences \citep{Lu01, Bakoyannis10}. Accordingly, valid statistical inference in the presence of missing causes of failure requires methods that appropriately account for missingness. Knowing how robust the findings are under different missing data mechanisms becomes a question of great practical importance.

An example arises from the East Africa Regional Consortium of the International epidemiology Databases to Evaluate AIDS (EA-IeDEA), which collects data from HIV care programs in Kenya, Uganda, and Tanzania. The consortium aims to identify factors associated with gaps in care and mortality following the initiation of antiretroviral therapy (ART). In this setting, under-reporting of death is common, leading to misclassified outcomes. Specifically, patients who die outside the care system may be incorrectly recorded as lost to follow-up, creating ambiguity about their true cause of failure. To mitigate this problem, EA-IeDEA conducts community tracing of patients lost to care to verify their vital status. However, even with extensive tracing, many individuals cannot be located, and those successfully traced are not necessarily representative of all lost patients \citep{Brinkhof09, Brinkhof10}. Non-traced individuals may be more likely to have died than those who have been successfully traced \citep{Brinkhof09, Brinkhof10}. Consequently, the probability that the cause of failure is missing depends on the unobserved event type itself, thus violating the key missing at random (MAR) assumption and motivating the current research for assessing the robustness of the study findings under missing-not-at-random (MNAR) mechanisms.

The issue of missing causes of failure under MAR has received considerable attention in the literature of competing risks analysis. Nonparametric estimation and two-sample hypothesis testing for the cumulative incidence function under MAR cause of failure has been addressed using multiple imputation \citep{Lee11, Lee14} and nonparametric pseudolikelihood approaches \citep{Bakoyannis19, Bakoyannis20B}. Several papers have addressed the MAR cause of failure problem in the context of modeling the cause-specific hazard function \citep{Goetghebeur95, Lu01, Gao05, Hyun12, Bordes14, Nevo17, Bakoyannis20, Lo22, Zhou23} and the cumulative incidence function \citep{Bakoyannis10,Moreno13, Mao17}. Violation of the MAR assumption, which cannot be verified based on the observed data, can invalidate the latter methods. Nevertheless, the problem of regression analysis of competing risks data with MNAR cause of failure has not received sufficient attention. 
\citet{Moreno15} proposed a sensitivity analysis approach for MNAR cause of failure based on a pattern-mixture model specification. \citet{Azizi20} proposed a Bayesian sensitivity analysis approach for evaluating the robustness of Bayesian estimates to violations of the assumption of MAR cause of failure. However, existing sensitivity analysis approaches focus on point estimates only and do not consider the additional uncertainty due to the unknown MNAR scenario. Thus, they cannot be used to directly assess the impact of violations of the MAR assumption on hypothesis testing and the associated conclusions. Furthermore, these approaches use fully parametric competing risks models, which limits their applicability. In addition to these sensitivity analysis approaches, \citet{Koley22} proposed a methodology for regression analysis with current status competing risks data and MNAR cause of failure. However, to achieve identifiability, this approach imposes strong and untestable assumptions based on the data, and also uses a fully parametric competing risks model. 

To assess the robustness of study findings, a number of sensitivity analysis approaches have been proposed for general MNAR and unmeasured confounding problems \citep{Vansteelandt06, Todem10, Cao13, Zhao19, Ding16, Vanderweele17, Bonvini22}. However, these methods are not applicable to the problem of quantifying robustness of competing risks analyses when the causes of failure are MNAR. A review of these approaches in relation to our proposed methodology is provided in the Discussion Section.

In this paper, we propose a principled sensitivity analysis for the semiparametric proportional cause-specific hazards model. The approach constructs a class of MNAR mechanisms, including MAR as a special case, under a pattern-mixture model specification. The goal is to estimate functional regression coefficients as functions of a sensitivity parameter, which controls the extent of deviation from the MAR assumption. The novelty of this work is twofold. First, unlike previous sensitivity analysis approaches for this problem \citep{Moreno15}, we account for the additional uncertainty regarding the unknown MNAR scenario by proposing a simultaneous confidence band for the functional regression coefficients. Given that these are infinite-dimensional parameters, we use empirical process theory~\citep{Kosorok08} to study the methodology for deriving this confidence band. Second, we introduce the concept and methodology for computing a \textit{robustness interval}, which is defined as the range of MNAR scenarios (i.e., sensitivity parameter values) within which an effect remains statistically significant. This interval provides a measure of the extent of the robustness of conclusions about statistical hypotheses of interest against MNAR causes.

The rest of the paper is organized as follows: We introduce notation, the proposed methodology, and cover the relevant asymptotic theory in Section~\ref{s:methods}. In addition to establishing the asymptotic properties of the proposed estimator, we evaluate the finite-sample performance of our methodology through extensive simulation experiments. These are presented in Section~\ref{s:sims}. We illustrate the proposed methodology in a real-world setting by using competing risk data from the EA-IeDEA study cohort. This analysis, which is presented in Section~\ref{s:analysis}, indicates that our previously published results regarding key factors associated with the hazard of gap in care are quite robust against violations of the MAR assumption.  We summarize our methodology and connect our ideas with other sensitivity analysis approaches for general MNAR and unmeasured confounding problems in a brief discussion in Section~\ref{s:discussion}.

\section{Methodology \label{s:methods}}
\subsection{Notation and data}
Consider a cohort study with finite follow-up duration $\tau$. Let $T$ denote the failure time, $U$ the right censoring time, $X=T\wedge U$ the observation time, where $a\wedge b = \min(a,b)$, and $\Delta = I(T\leq U)$ the event indicator. Denoting the cause of failure by $C$, we assume, for simplicity and without loss of generality, that $C\in\{1,2\}$, although this methodology can be trivially adapted for situations with more than two causes of failure. In our motivating application, $C=1$ denotes gap in care and $C=2$ death. In addition, let $Z\in \mathcal{Z}\subset\mathbb{R}^p$ be the vector of covariates of scientific interest. In a setting with missing causes of failure, we introduce the \textit{observed cause} indicator $R$, with $R=1$ if the cause of failure is observed and $R=0$ otherwise. In our setting the censoring status is always observed, so that if $R=0$, then $\Delta=1$. In some settings, an auxiliary covariate vector $A\in\mathcal{A}\subset\mathbb{R}^q$, which includes covariates that may be related to the probability of missingness, is also observed. The observable data from a study of $n$ individuals are independent and identically distributed copies of
\[
D_i=\left\{\begin{array}{cc}
(X_i,\Delta_i,C_i, Z_i, A_i, R_i) & \mbox{if} \ \ R_i=1 \ \ \textrm{and} \ \ \Delta_i=1 \\
(X_i,\Delta_i,Z_i,A_i,R_i) & \mbox{otherwise}
\end{array}, \right. \ \ i=1,\ldots,n,
\]
where the cause of failure is only observed for those with $\Delta_i=1$ and $R_i=1$. To facilitate the presentation of the methods, we introduce some further notation. Let $\Delta_{ij}=I(\Delta_i=1,C_i=j)$, $j=1,2$, be the indicator that the $i$th individual experienced the $j$th cause of failure. In addition, we define the counting processes $N_i(t)=I(X_i\leq t, \Delta_i=1)$, $t\in[0,\tau]$, and the cause-specific counting process $N_{ij}(t)=I(X_i\leq t, \Delta_{ij}=1)=\Delta_{ij}N_i(t)$, $j=1,2$. Finally, we define the at-risk process $Y_i(t)=I(X_i\geq t)$, $t\in[0,\tau]$.

\subsection{Competing risk analysis with MAR cause of failure}

In this work, we focus on the cause-specific hazard function, which is defined, conditionally on covariates $Z$, as 
\[
\lambda_j(t;z)=\lim_{h\downarrow 0}\frac{P(t\leq T<t+h, C=j|T\geq t, Z=z)}{h}, \qquad t\in[0,\tau], \ \ j=1,2.
\]
We impose the semiparametric proportional hazards model
\begin{equation}
\lambda_j(t;z)=\lambda_{0,j}(t)\exp(\beta_{0,j}^{\prime}z) \qquad t\in[0,\tau], \ \ j=1,2, \label{model}
\end{equation}
where $\lambda_{0,j}(t)$ is the unspecified baseline cause-specific hazard corresponding to the $j$th cause of failure. In the absence of missingness, the parameters in model \eqref{model} can be estimated by maximizing the logarithm of the partial likelihood function
\[
l_n(\beta) = \sum_{j=1}^2\sum_{i=1}^n\int_0^{\tau}\left[\beta_j^{\prime} Z_i - \log\left\{\sum_{l=1}^nY_l(t)\exp(\beta_j^{\prime} Z_l)\right\}\right]\textrm{d}N_{ij}(t), 
\]
where $\beta = (\beta_1^{\prime},\beta_2^{\prime})^{\prime}$, or equivalently, by solving the partial score equation 
\[
G_{n,j}(\beta_j)=\sum_{i=1}^n\int_0^{\tau}\left\{Z_i - \frac{\sum_{l=1}^nZ_lY_l(t)e^{\beta_j^{\prime} Z_l}}{\sum_{l=1}^nY_l(t)e^{\beta_j^{\prime} Z_l}}\right\}\textrm{d}N_{ij}(t)=0, \qquad j=1,2.
\]
When there are missing causes of failure however, $G_{n,j}(\beta_j)$, $j=1,2$, cannot be computed. If cause of failure is MAR conditionally on the fully observed variables, $W=(X,Z^{\prime},A^{\prime})^{\prime}$ and $\Delta=1$, then
\begin{equation}
    P(C=2|R=0,\Delta=1, W) = P(C=2|R=1,\Delta=1, W). \label{MAR}
\end{equation}
In other words, the model for the probability of the cause of failure for the missing cases is
the same as that for the observed cases. For simplicity and following \citet{Bakoyannis20}, we impose a parametric logit model $P(C=2|R=1,\Delta=1, W)=g(\gamma_0^{\prime}\tilde{W})$, where $g(x)=e^x/(1+e^x)$, $\gamma_0\in\Gamma\subset\mathbb{R}^{p+q +2}$, and $\tilde{W}=(1,W^{\prime})^{\prime}$. The proposed approach is trivially applicable with a broad class of models for $P(C=2|R=1,\Delta=1, W)$, for example, a flexible logistic regression model that incorporates regression splines for the continuous variables \citep{Wu25}. The challenge arises in cases where the MAR assumption does not hold, since the equality \eqref{MAR} is no longer valid and thus the model $g(\gamma_0^{\prime}\tilde{W})$, which is established on individuals with observed causes of failure, cannot be used for predicting the missing causes. Consequently, under MNAR, methods that assume that cause of failure is MAR are expected to lead to biased estimates. 

\subsection{Identification region under MNAR \label{ss:IR}}
In this subsection, we present a sensitivity analysis approach under MNAR cause of failure using a sensitivity parameter which quantifies the extent to which the MAR assumption is violated. This parameter can receive a pre-specified range of values, which can be elicited from subject-matter experts or study investigators. Under this range, a plausible set of regression coefficients can be estimated.

The proposed approach assumes a class of MNAR mechanisms under a pattern-mixture model specification. A natural choice for such a class is
\[
\mathcal{M}=\big\{P(C=2|R=0,\Delta=1,W)=g\{\gamma_0^{\prime}\tilde{W}+s(\eta,W)\}:\eta\in\mathcal{H}\subset\mathbb{R}^r\big\},
\]
for a finite $r$. In this framework, the departure from MAR is quantified by the term $s(\eta,W)$. This term depends on the fully observed variables $W$ and the non-estimable \textit{sensitivity} parameter $\eta$. 
For the remainder of this paper, to simplify the presentation of the methodology and facilitate interpretability, we focus on the special subset of this class
\[
\tilde{\mathcal{M}}=\big\{P(C=2|R=0,\Delta=1,W)=g(\gamma_0^{\prime}\tilde{W}+\eta):\eta\in[a,b]\subset\mathbb{R}\big\}\subset \mathcal{M}.
\]
Under this subclass $\tilde{\mathcal{M}}$, $\eta=0$ corresponds to the MAR mechanism, while $\eta\neq 0$ implies MNAR. In our motivating study, $\eta$ is the log odds ratio of death for those with missing vital status versus those with observed vital status, so that
\[
\exp(\eta) = \frac{P(C=2|R=0,\Delta=1,W)}{1-P(C=2|R=0,\Delta=1,W)}\bigg/ \frac{P(C=2|R=1,\Delta=1,W)}{1-P(C=2|R=1,\Delta=1,W)}.
\]
The sensitivity parameter $\eta$ takes values in a set $[a,b]$, which is selected by subject-matter experts or study investigators. In the next subsection we consider an approach that is useful in situations where obtaining a plausible range of values for $\eta$ is not possible or desirable. 

Given that the cause of failure $C$ is not observed for the missing cases (i.e., observations with $R=0$ and $\Delta=1$), $\eta$ cannot be estimated from the data. For given $\eta$ and the parameter $\gamma_0$, which is estimable from the complete data, we define the expected jump processes
\[
\tilde{N}_{2}(t;\gamma_0, \eta) = \{R\Delta_{2} + (1 - R)g(\gamma_0^{\prime}\tilde{W} + \eta)\}N(t)
\]
and $
\tilde{N}_{1}(t;\gamma_0, \eta) = [R\Delta_{1} + (1 - R)\{1 - g(\gamma_0^{\prime}\tilde{W} + \eta)\}]N(t)$, $t\in[0,\tau]$.
If the cause of failure is missing, the size of the jump of these processes is equal to the model-based probabilities of death $g(\gamma_0^{\prime}W + \eta)$ and gap in care $1 - g(\gamma_0^{\prime}W + \eta)$, respectively. Otherwise, these jump sizes are equal to 1 or 0, corresponding to the values of $\Delta_{2}$ and $\Delta_{1}$.

For a given value of $\eta$ under the class $\tilde{\mathcal{M}}$ (i.e., for a given MNAR mechanism), the population regression parameter, denoted by $\beta_j^*(\eta)$, $j=1,2$, is the root of the equation
\[
\tilde{G}_{j}^*(\beta_j^*(\eta);\gamma_0,\eta)=E\left(\int_0^{\tau}\left[Z - \frac{E\{ZY(t)e^{\beta_j^*(\eta)^{\prime}Z}\}}{E\{Y(t)e^{\beta_j^*(\eta)^{\prime}Z}\}}\right]\textrm{d}\tilde{N}_{j}(t;\gamma_0,\eta)\right)=0,  \qquad j=1,2.
\]
The corresponding set of population parameters under the entire class of MNAR mechanisms $\tilde{\mathcal{M}}$ is $
B_j^*\equiv \{\beta_j^*(\eta):\eta\in[a,b]\}$, $j=1,2$.
Such sets are known as \textit{ignorance regions} \citep{Vansteelandt06} or \textit{identification regions} \citep{Bonvini22} in the general sensitivity analysis literature. In this paper, we use the latter term. The functional regression coefficients can be consistently estimated using our proposed functional partial pseudo-score estimating equations
\begin{equation}
\tilde{G}_{n,j}(\hat{\beta}_{n,j}(\eta);\hat{\gamma}_n,\eta)=\frac{1}{n}\sum_{i=1}^n\int_0^{\tau}\left\{Z_i - \frac{\sum_{l=1}^nZ_lY_l(t)e^{\hat{\beta}_{n,j}(\eta)^{\prime} Z_l}}{\sum_{l=1}^nY_l(t)e^{\hat{\beta}_{n,j}(\eta)^{\prime} Z_l}}\right\}\textrm{d}\tilde{N}_{ij}(t;\hat{\gamma}_n,\eta)=0, \label{estfun}   
\end{equation}
for $j=1,2$ and $\eta\in[a,b]$, where $\hat{\gamma}_n$ is the maximum likelihood estimate of $\gamma_0$ based on the complete cases. The corresponding set of regression coefficient estimates under the entire class of MNAR mechanisms $\tilde{\mathcal{M}}$ is
\[
B_{n,j}\equiv\{\hat{\beta}_{n,j}(\eta):\eta\in[a,b]\}, \qquad j=1,2.
\]
Computation of $\hat{\beta}_{n,j}(\eta)$, $\eta\in[a,b]$, can be practically implemented in two stages. In the first stage, equation \eqref{estfun} is solved for a finite subset $\{a,\eta_1,\ldots,\eta_M,b\}$ of the sensitivity parameter set $[a,b]$, for a sufficiently large $M$. In the second stage, computation of $\hat{\beta}_{n,j}(\eta)$, for any $\eta\in[a,b]$, can be performed using linear or polynomial interpolation based on the values $\hat{\beta}_{n,j}(a),\hat{\beta}_{n,j}(\eta_1),\ldots,\hat{\beta}_{n,j}(\eta_M),\hat{\beta}_{n,j}(b)$. In Appendix A we argue that the function $\beta_j^*(\eta)$, $\eta\in[a,b]$, is continuous. 

Next, we construct a confidence band for a linear combination $K'\beta_j^*(\eta)$, for $K\in\mathbb{R}^p$, which can be directly applied to construct a confidence band for the $l$th functional regression coefficient $\beta_{j,l}^*(\eta)$, for $l=1,\ldots,p$, by choosing $K$ to be the indicator of the $l$th coordinate of $\beta_j^*(\eta)$. Let $\hat{\psi}_{ij}(\eta)$, $i=1,\ldots,n$, $j=1,2$, denote the empirical  version of the influence function ${\psi}_{ij}(\eta)$ of the proposed estimator (see Theorem 2 in Section~\ref{ss:theory} below and Appendix B for explicit formulas for $\hat{\psi}_{ij}(\eta)$). A simultaneous $100(1-\alpha)\%$ confidence band for $K'\beta^*_j(\eta)$, $\eta\in[a,b]$, can be computed based on the following wild bootstrap approach, which approximates the sampling distribution of the proposed estimator. Wild bootstrap avoids repeated model refitting while maintaining the correct covariance structure under complex censoring and missingness mechanisms, as justified by Theorem 3 in Section~\ref{ss:theory}.

Select a large integer $S$, and for $s\in\{1,\ldots,S\}$ follow the steps:
\vspace{7pt} \\
\indent \textit{Step} 1. Simulate i.i.d. random variables $\xi_1^{(s)},\ldots,\xi_n^{(s)}$ from $N(0,1)$.
\vspace{7pt} \\
\indent \textit{Step}  2. Based on $\xi_1^{(s)},\ldots,\xi_n^{(s)}$, calculate
\[
\hat{\mathbb{G}}_{n,j}^{(s)}(\eta)=\frac{1}{\sqrt{n}}\sum_{i=1}^n\hat{\psi}_{ij}(\eta)\xi_i^{(s)}, \ \ \eta\in[a,b],
\]
\indent and then $\sup_{\eta\in[a,b]}|K'\hat{\mathbb{G}}_{n,j}^{(s)}(\eta)|$.

When this process is completed, the empirical $100(1-\alpha)$th percentile of the sample 
\[
\sup_{\eta\in[a,b]}|K'\hat{\mathbb{G}}_{n,j}^{(1)}(\eta)|, \ldots,\sup_{\eta\in[a,b]}|K'\hat{\mathbb{G}}_{n,j}^{(S)}(\eta)|,
\]
denoted by $\hat{c}_{\alpha,j}$ is be computed. From this, the simultaneous confidence band is
\[
K'\hat{\beta}_{n,j}(\eta)\pm \frac{\hat{c}_{\alpha,j}}{\sqrt{n}}, \qquad \eta\in[a,b].
\]
Setting $S=1000$ was shown to perform well in simulation studies (Section~\ref{s:sims}). A conservative $100(1-\alpha)\%$ confidence interval for the identification region $\{K'\beta_j^*(\eta):\eta\in[a,b]\}$ can be computed as
\[
\left[\inf_{\eta\in[a,b]}\{K'\hat{\beta}_{n,j}(\eta)\}-\frac{\hat{c}_{\alpha,j}}{\sqrt{n}},\sup_{\eta\in[a,b]}\{K'\hat{\beta}_{n,j}(\eta)\}+\frac{\hat{c}_{\alpha,j}}{\sqrt{n}}\right].
\]
Given that the asymptotic coverage probability of the simultaneous confidence band is at least $1-\alpha$, as justified by Theorem 3 in Section~\ref{ss:theory}, the asymptotic coverage probability of the above confidence interval is at least $1-\alpha$.

\subsection{Robustness interval under MNAR}
\label{ss:sensitB}
In this subsection, we present an alternative sensitivity analysis approach for situations where a sensitivity parameter range $[a,b]$ cannot be prespecified. The main idea is to estimate, for all statistically significant effects in the MAR analysis, the largest absolute value of the sensitivity parameter (i.e., the most extreme violation of the MAR assumption) under which statistical significance is preserved. This maximal value is used to specify an entire range of MNAR mechanisms (i.e., sensitivity parameter values) within which statistical significance is preserved. We call this range a \textit{robustness interval}. 

For computational convenience, we consider symmetric (around 0) robustness intervals here. Before providing the approach for the computation of these intervals, we introduce some further notation. For any nonnegative real number $\bar{\eta}\geq 0$, let $\hat{c}_{\alpha,j}(\bar{\eta})$ denote the empirical $100(1-\alpha)$th percentile for the $100(1-\alpha)\%$ simultaneous confidence band for $K'\beta_j^*(\eta)$ for all $\eta\in[-\bar{\eta},\bar{\eta}]$, computed based on the wild bootstrap approach described above. Also, define the $100(1-\alpha)\%$ confidence interval
\[
\mathcal{I}_{n,j}(K;\bar{\eta},\alpha)\equiv \left[\inf_{\eta\in[-\bar{\eta},\bar{\eta}]}\{K'\hat{\beta}_{n,j}(\eta)\}-\frac{\hat{c}_{\alpha,j}(\bar{\eta})}{\sqrt{n}},\sup_{\eta\in[-\bar{\eta},\bar{\eta}]}\{K'\hat{\beta}_{n,j}(\eta)\}+\frac{\hat{c}_{\alpha,j}(\bar{\eta})}{\sqrt{n}}\right],
\]
and the corresponding identification region $B_j^*(K;\bar{\eta})\equiv\{K'\beta_j^*(\eta):\eta\in[-\bar{\eta},\bar{\eta}]\}$. Note that, both $\mathcal{I}_{n,j}(K;\bar{\eta},\alpha)$ and $B_j^*(K;\bar{\eta})$ are monotonic in $\bar{\eta}$ in the sense that if $\eta_1\leq\eta_2$, then $\mathcal{I}_{n,j}(K;\eta_1,\alpha)\subset \mathcal{I}_{n,j}(K;\eta_2,\alpha)$, for any sample size $n$, and $B_j^*(K;\eta_1)\subset B_j^*(K;\eta_2)$. Our goal is to compute the largest value $\tilde{\eta}_n\geq 0$, such that the $100(1-\alpha)\%$ confidence interval $\mathcal{I}_{n,j}(K;\tilde{\eta}_n,\alpha)$
for the corresponding identification region $B_j^*(K;\tilde{\eta}_n)$ does not include zero. Thus, $\tilde{\eta}_n$ is defined as
\[
\tilde{\eta}_n=\sup\left\{\bar{\eta}:\bar{\eta}\geq 0,\left[\inf_{\eta\in[-\bar{\eta},\bar{\eta}]}\{K'\hat{\beta}_{n,j}(\eta)\}-\frac{\hat{c}_{\alpha,j}(\bar{\eta})}{\sqrt{n}}\right]\times\left[\sup_{\eta\in[-\bar{\eta},\bar{\eta}]}\{K'\hat{\beta}_{n,j}(\eta)\}+\frac{\hat{c}_{\alpha,j}(\bar{\eta})}{\sqrt{n}}\right]\geq \epsilon\right\},
\]
for a small $\epsilon>0$. A positive value for the product of the two factors in brackets above indicates that the limits of the $100(1-\alpha)\%$ confidence interval have the same sign and, thus, the confidence interval does not include zero. The robustness interval is $[-\tilde{\eta}_n,\tilde{\eta}_n]$, in the scale of $\eta$, or, equivalently, $[e^{-\tilde{\eta}_n},e^{\tilde{\eta}_n}]$, in the odds ratio scale. This interval is the range of MNAR scenarios within which statistical significance is preserved, based on the observed data $D_1,\ldots,D_n$. To see the validity of this interpretation, suppose that $0\in B_j^*(K;\eta_1)$, for some $\eta_1\in[0,\tilde{\eta}_n]$. This implies that, $0\in B_j^*(K;\bar{\eta})$ for all $\bar{\eta}\geq \eta_1$, by the monotonicity of $B_j^*(K;\bar{\eta})$ in $\bar{\eta}$. Then, the probability of a Type I error is
\begin{eqnarray*}
   P\left(\cup_{\bar{\eta}\in [\eta_1,\tilde{\eta}_n]}\left\{0\notin\mathcal{I}_{n,j}(K;\bar{\eta},\alpha)\right\}\right) &=& P\left(0\notin\mathcal{I}_{n,j}(K;\eta_1,\alpha)\right) \\
   &=& 1 -  P\left(0\in\mathcal{I}_{n,j}(K;\eta_1,\alpha)\right) \\
   &\leq& 1 - P\left(B_j^*(K;\eta_1)\subset\mathcal{I}_{n,j}(K;\eta_1,\alpha)\right) \\
   &\leq& \alpha + o(1),
\end{eqnarray*}
where the first equality follows from the monotonicity of $\mathcal{I}_{n,j}(K;\bar{\eta},\alpha)$ in $\bar{\eta}$, and the last inequality from Theorem 3 below, which implies that
\begin{eqnarray*}
1 - \alpha + o(1)&\leq& P\left(K'\hat{\beta}_{n,j}(\eta)-\frac{\hat{c}_{\alpha,j}(\eta_1)}{\sqrt{n}}\leq K'\beta_j^*(\eta)\right. \\
&& \hspace{15mm}\left. \leq K'\hat{\beta}_{n,j}(\eta)+\frac{\hat{c}_{\alpha,j}(\eta_1)}{\sqrt{n}} \ \ \textrm{for all} \ \ \eta\in[-\eta_1,\eta_1]\right) \\
&\leq & P\left(B_j^*(K;\eta_1)\subset\mathcal{I}_{n,j}(K;\eta_1,\alpha)\right).
\end{eqnarray*}
Thus, the asymptotic Type I error is at most $\alpha$. The computation of $\tilde{\eta}_n$ can be achieved using a non-derivative-based root-finding algorithm to find the largest root of the equation
\[
f_{n,j}(\bar{\eta};K,\epsilon)\equiv\left[\inf_{\eta\in[-\bar{\eta},\bar{\eta}]}\{K'\hat{\beta}_{n,j}(\eta)\}-\frac{\hat{c}_{\alpha,j}(\bar{\eta})}{\sqrt{n}}\right]\times\left[\sup_{\eta\in[-\bar{\eta},\bar{\eta}]}\{K'\hat{\beta}_{n,j}(\eta)\}+\frac{\hat{c}_{\alpha,j}(\bar{\eta})}{\sqrt{n}}\right] - \epsilon = 0
\]
with respect to $\bar{\eta}$ within a search set $[0,\eta_{\max}]$, $\eta_{\max}>0$, given a small value for $\epsilon$. The applicability of these algorithms to this problem is guaranteed by the fact that the function $f_{n,j}(\bar{\eta};K,\epsilon)$ is continuous in $\bar{\eta}$, by the continuity of the proposed estimator in the sensitivity parameter $\eta$. In the application, we search for roots in the set $[0,\eta_{\max}]=[0,5]$, which covers a very broad range of MNAR scenarios, with $\epsilon = 10^{-8}$, using the R function \texttt{uniroot.all}. If $f_{n,j}(0;K,\epsilon)\times f_{n,j}(\eta_{\max};K,\epsilon)< 0$, that is, if there is a sign change in $f_{n,j}(\bar{\eta};K,\epsilon)$ on the search set $[0,\eta_{\max}]$, then there exists at least one root in $[0,\eta_{\max}]$, by the intermediate value theorem. Otherwise, if $f_{n,j}(0;K,\epsilon)\times f_{n,j}(\eta_{\max};K,\epsilon)> 0$, there are no roots in $[0,\eta_{\max}]$ since there can only be one sign change in $f_{n,j}(\bar{\eta};K,\epsilon)$, from positive to negative, on any search set. This is because once the confidence interval captures zero, i.e., if $0\in\mathcal{I}_{n,j}(K;\eta_1,\alpha)$ for some $\eta_1>0$ and thus $f_{n,j}(\eta_1;K,\epsilon)<0$, then zero will be included in any other confidence interval over a wider sensitivity parameter range, i.e. $0\in\mathcal{I}_{n,j}(K;\eta_2,\alpha)$ and $f_{n,j}(\eta_2;K,\epsilon)<0$ for all $\eta_2\geq\eta_1$. Therefore, when there are no roots in $[0,\eta_{\max}]$, the robustness interval is defined either as the empty set $\emptyset$, if $f_{n,j}(0;K,\epsilon)<0$ (i.e., effect is not significant in the MAR analysis), or equal to $[-\eta_{\max},\eta_{\max}]$ (or $[e^{-\eta_{\max}},e^{\eta_{\max}}]$ in the odds ratio scale), if $f_{n,j}(\eta_{\max};K,\epsilon)> 0$ (i.e., effect remains significant over the entire search set $[-\eta_{\max},\eta_{\max}]$).

A na\"ive approach to estimating robustness intervals is to compute the $100(1-\alpha)\%$ confidence band over the broadest set $[-\eta_{\max},\eta_{\max}]$ and identify the range of $\eta$ values where the band excludes zero. However, this approach is conservative, tending to yield unnecessarily narrower robustness intervals because, by definition, $\hat{c}_{\alpha,j}(\eta_{\max})\geq \hat{c}_{\alpha,j}(\bar{\eta})$, for any $\bar{\eta}\in[0,\eta_{\max}]$, and thus
\[
\left[\inf_{\eta\in[-\bar{\eta},\bar{\eta}]}\{K'\hat{\beta}_{n,j}(\eta)\}-\frac{\hat{c}_{\alpha,j}(\eta_{\max})}{\sqrt{n}},\sup_{\eta\in[-\bar{\eta},\bar{\eta}]}\{K'\hat{\beta}_{n,j}(\eta)\}+\frac{\hat{c}_{\alpha,j}(\eta_{\max})}{\sqrt{n}}\right] \supset \mathcal{I}_{n,j}(K;\bar{\eta},\alpha),
\]
for all $\bar{\eta}\in[0,\eta_{\max}]$. Our proposed method yields wider robustness intervals while controlling the Type I error in the strong sense asymptotically, as shown previously. The inferior performance of the na\"ive approach for computing robustness intervals is illustrated in our real data analysis (Section~\ref{s:analysis})

\subsection{Asymptotic theory \label{ss:theory}}
The theorems presented in this subsection guarantee the validity of the proposed methodology under a set of realistic regularity conditions. These conditions along with the proofs of the following theorems are listed in Appendix A. Before presenting Theorem 1, consider the notation $\overset{as*}\rightarrow$ to indicate outer almost sure convergence \citep{Kosorok08}. Let $\hat{\beta}_{n,j}(\cdot)$ and $\beta_j^*(\cdot)$ denote the functional regression coefficients $\hat{\beta}_{n,j}(\eta)$ and $\beta_j^*(\eta)$, $\eta\in[a,b]$, as whole functions as opposed to their values for a particular choice of $\eta$. Theorem 1 below justifies the validity of the proposed estimator of the functional regression coefficient.
\begin{theorem}
Assume that regularity conditions C1--C6 listed in Appendix A hold. Then, 
\[
\sup_{\eta\in[a,b]}\left\|\hat{\beta}_{n,j}(\eta)-\beta_j^*(\eta)\right\|\overset{as*}\rightarrow 0, \ \ j=1,2.
\]
Moreover, if the assumed class $\tilde{\mathcal{M}}$ includes the true probability $P(C=2|R=0,\Delta=1,W=w)$, then
\[
\inf_{\eta\in[a,b]}\left\|\hat{\beta}_{n,j}(\eta)-\beta_{0,j}\right\|\overset{as*}\rightarrow 0, \ \ j=1,2.
\]
\end{theorem}
Theorem 1 states that the proposed estimator $\hat{\beta}_{n,j}(\cdot)$ is strongly uniformly consistent for the parameter $\beta_j^*(\cdot)$, $j=1,2$. This justifies the use of the proposed estimator for sensitivity analysis purposes. The second statement implies that the sets $B_{n,j}$ will contain the true parameters $\beta_{0,j}$, $j=1,2$, almost surely asymptotically.

The next theorem provides the asymptotic distribution of the estimator $\hat{\beta}_{n,j}(\cdot)$. The influence function of this estimator, denoted by $\psi_{ij}(\eta)$, $i=1,\ldots,n$, $j=1,2$ and $\eta\in[a,b]$, is provided in Appendix A.
\begin{theorem}
    Assume that regularity conditions C1--C6 listed in Appendix A hold. Then,
    \[
    \sqrt{n}\left\{\hat{\beta}_{n,j}(\eta)-\beta_j^*(\eta)\right\}=\frac{1}{\sqrt{n}}\sum_{i=1}^n\psi_{ij}(\eta)+\epsilon_{n,j}(\eta), \ \ j=1,2, \ \ \eta\in[a,b]
    \]
    where $E\{\psi_j(\eta)\}=0$ and $E\|\psi_j(\eta)\|^2<\infty$, $\eta\in[a,b]$, and $\sup_{\eta\in[a,b]}\|\epsilon_{n,j}(\eta)\|=o_p(1)$, $j=1,2$. Moreover, the classes of influence functions $\{\psi_j(\eta):\eta\in[a,b]\}$, $j=1,2$, are Donsker.
\end{theorem}
Theorem 2 implies that $\sqrt{n}\{\hat{\beta}_{n,j}(\cdot)-\beta_j^*(\cdot)\}$ converges weakly to a (vector-valued) tight zero mean Gaussian process $\mathbb{G}_j(\cdot)$, with covariance function $\sigma_{j}(\eta_1,\eta_2)=E\{\psi_j(\eta_1)\psi_j'(\eta_2)\}$. This covariance can be consistently estimated by $\hat{\sigma}_{n,j}(\eta_1,\eta_2)=n^{-1}\sum_{i=1}^n\hat{\psi}_{ij}(\eta_1)\hat{\psi}_{ij}'(\eta_2)$, where $\hat{\psi}_{ij}$ are the empirical versions of the influence functions. Explicit formulas for these are provided in Appendix B.

Next, define the multiplier empirical process $\hat{\mathbb{G}}_{n,j}(\eta)=n^{-1/2}\sum_{i=1}^n\hat{\psi}_{ij}(\eta)\xi_i$, where $\xi_i$ are standard normal variables that are independent of the data $\mathcal{D}_n=(D_1,\ldots,D_n)$. Theorem 3 justifies the validity of the proposed wild bootstrap approach for computing simultaneous confidence bands for any linear combination $K'\beta^*_j(\eta)$, $\eta\in[a,b]$. Therefore, it also justifies the confidence intervals for the identification region and the proposed robustness intervals.
\begin{theorem}
    Assume that regularity conditions C1--C6 listed in Appendix A hold. Then, for any fixed vector $K\in\mathbb{R}^p$, we have that
    \[
    \sup_{x\in A}\left|P\left\{\sup_{\eta\in[a,b]}|K'\hat{\mathbb{G}}_{n,j}(\eta)|\leq x\bigg|\mathcal{D}_n\right\}-P\left\{\sup_{\eta\in[a,b]}|K'\mathbb{G}_j(\eta)|\leq x\right\}\right|\overset{p}\rightarrow 0, \qquad j=1,2,
    \]
    for any closed set $A\subset\mathbb{R}\cup\{-\infty,\infty\}$.
\end{theorem}

\section{Simulation Studies \label{s:sims}}

To evaluate the finite-sample performance of the proposed approach, we conducted a series of simulation experiments. We considered two causes of failure, with $C\in\{1,2\}$, and one covariate $Z\sim N(0,1)$. The competing risks data were simulated based on the cause-specific hazard functions $\lambda_j(t;Z)=p_0\lambda_0^{p_0}t^{p_0-1}\exp(\beta_{0,j}Z)$, $j=1,2$, where $p_0=1.5$, $\lambda_0=1.5$, $\beta_{0,1}=0.5$, and $\beta_{0,2}=-1$, while the right censoring time $U$ was simulated from $\textrm{Exp}(0.7)$. This simulation setup led to an average proportion of $C=1$ among the non-right-censored observations of 49.0\%, with the average right-censoring rate being $20.5\%$. The nonmissingness indicators $R$ were simulated according to the following four MNAR scenarios:
\begin{enumerate}
    \item $\textrm{logit}\{P(R=1|\Delta=1,X,Z,C)\}=0.3+0.5\times I(C=2)$. 
    \item $\textrm{logit}\{P(R=1|\Delta=1,X,Z,C)\}=0.3+I(C=2)$. 
    \item $\textrm{logit}\{P(R=1|\Delta=1,X,Z,C)\}=0.3-X+Z+0.5\times I(C=2)$.
    \item $\textrm{logit}\{P(R=1|\Delta=1,X,Z,C)\}=0.3-X+Z+I(C=2)$.
\end{enumerate}
These scenarios led to average missing cause of failure rates of 48.6\%, 54.2\%, 56.3\%, and 61.4\%, respectively. For each scenario, we simulated 1,000 datasets and considered the sample sizes $n=200, 400, 800$. The data were analyzed based on the partial pseudolikelihood approach by \citet{Bakoyannis20}, which assumes MAR, and the proposed sensitivity analysis methodology. For the proposed methodology, we assumed the pattern-mixture model 
\[
\textrm{logit}\{P(C=2|\Delta=1,X,Z,R=0)\}=\gamma_0'\tilde{W}+\eta, \qquad \eta\in[-1,1],
\]
where $\tilde{W}=(1,X,Z)'$. Under this simulation setup, the corresponding true model is
\[
\textrm{logit}\{P(C=2|\Delta=1,X,Z,R=0)\}=0.5Z+\log\left\{\frac{1-P(R=1|\Delta=1,X,Z,C=2)}{1-P(R=1|\Delta=1,X,Z,C=1)}\right\}+\eta_0,
\]
where $\eta_0\in\{0.5,1\}$, depending on the scenario. Thus, our assumed pattern-mixture model was correctly specified under missingness scenarios 1 and 2, but misspecified in scenarios 3 and 4. A similar model was considered in the implementation of the approach by \citet{Bakoyannis20}, with the exception that $\eta=0$, which corresponds to the MAR scenario. The evaluation metrics in this simulation study were the pointwise bias $1000^{-1}\sum_{s=1}^{1000}\hat{\beta}_{n,1,s}(\eta)- \beta_{1}^*(\eta)$, $\eta\in\{-1,-0.5,0,0.5,1\}$, where the subindex $s$ indicates the simulated dataset number, the mean minimum absolute deviation, defined as $1000^{-1}\sum_{s=1}^{1000}\inf_{\eta\in[-1,1]}|\hat{\beta}_{n,1,s}(\eta)-\beta_{0,1}|$, and the empirical coverage probabilities of the proposed 95\% simultaneous confidence band for $\beta_{1}^*(\eta)$, $\eta\in[-1,1]$, and the proposed 95\% confidence interval for the identification region. For comparison, we also computed the bias and 95\% confidence interval for the true $\beta_{0,1}$ under the MAR assumption. 

The results from the simulation study are presented in Table~\ref{t:sims}. In all scenarios, the pointwise bias was close to zero and the mean minimum absolute deviation was small and decreased with sample size $n$. This provides empirical evidence for the validity of the proposed estimator and its robustness against some degree of misspecification of the pattern-mixture model $\textrm{logit}\{P(C=2|\Delta=1,X,Z,R=0)\}$ (scenarios 3 and 4). The empirical coverage probabilities for the proposed 95\% simultaneous confidence bands and the 95\% confidence intervals for the identification region were close to the nominal level, regardless of the correctness of the specification of the pattern-mixture model. An exception was a somewhat lower coverage probability for the band when $n=200$, the smallest sample size considered here (empirical coverage probability range: 0.919 to 0.929). This was also observed for the confidence interval of the identification region in scenario 4 (empirical coverage probability: 0.926). This is expected to be due to the smaller sample size in addition to the large proportion of missingness (range of average missingness rate: 48.6\% to 61.4\%). The estimator that assumes MAR \citep{Bakoyannis20} exhibited large bias (bias range: 13.8\% to 48.6\%) and the corresponding 95\% confidence intervals provided poor coverage (empirical coverage probability range: 0.386 to 0.908). This suboptimal performance was more pronounced with higher missingness rates, more extreme violation of the MAR assumption (scenarios 2 and 4 where $\eta_0=1$), and larger sample sizes $n$. The last trend is attributed to the fact that the width of the confidence interval reduces with sample size, and also the inconsistency of the estimator due to the violation of the MAR assumption. These result in a coverage probability going to zero as $n\rightarrow\infty$.

\begin{table}
\caption{Results from the simulation study. (Bias$^*(\eta)$: Pointwise bias $1000^{-1}\sum_{s=1}^{1000}\hat{\beta}_{n,1,s}(\eta) - \beta_1^*(\eta)$, $\eta\in\{-1,-0.5,0,0.5,1\}$; MAD: mean minimum absolute deviation defined as $1000^{-1}\sum_{s=1}^{1000}\inf_{\eta\in[-1,1]}|\hat{\beta}_{n,1,s}(\eta)-\beta_{0,1}|$; CP$_{\textrm{Band}}$: Coverage probability of the 95\% simultaneous confidence band for $\beta_{1}^*(\eta)$, $\eta\in[-1,1]$,; CP$_{\textrm{IR}}$: Coverage probability of the 95\% confidence interval for the identification region; Bias: Bias of the estimate of $\beta_{0,1}$ assuming MAR; CP: Coverage probability of the 95\% confidence interval for $\beta_{0,1}$ assuming MAR)}
\label{t:sims}
\begin{center}
\begin{tabular}{cccccccccccc}
\hline
  & & \multicolumn{8}{c}{Proposed Approach} & \multicolumn{2}{c}{MAR} \\
  \cmidrule(lr){3-10} \cmidrule(lr){11-12}
  & & \multicolumn{5}{c}{Bias$^*(\eta)$} &&&&& \\
  \cmidrule(lr){3-7} 
 Scenario & $n$ & -1 & 0.5 & 0 & 0.5 & 1 & MAD & CP$_{\textrm{Band}}$ & CP$_{\textrm{IR}}$ & Bias & CP \\
 \hline
1 & 200 & 0.006 & 0.007 & 0.009 & 0.011 & 0.013 & 0.057 & 0.928 & 0.942 & 0.077 & 0.908 \\
 & 400 & 0.000 & 0.000 & 0.001 & 0.002 & 0.003 & 0.027 & 0.938 & 0.945 & 0.070 & 0.897 \\
 & 800 & 0.000 & 0.000 & 0.001 & 0.001 & 0.002 & 0.011 & 0.959 & 0.971 & 0.069 & 0.883 \\[5pt]
2 & 200 & 0.012 & 0.013 & 0.015 & 0.018 & 0.021 & 0.093 & 0.929 & 0.944 & 0.165 & 0.834 \\
 & 400 & 0.002 & 0.002 & 0.003 & 0.004 & 0.005 & 0.055 & 0.942 & 0.955 & 0.153 & 0.761 \\
 & 800 & 0.001 & 0.001 & 0.001 & 0.002 & 0.003 & 0.035 & 0.960 & 0.971 & 0.151 & 0.608 \\[5pt]
3 & 200 & 0.011 & 0.013 & 0.015 & 0.019 & 0.024 & 0.054 & 0.923 & 0.934 & 0.121 & 0.872 \\
 & 400 & -0.001 & 0.000 & 0.002 & 0.005 & 0.008 & 0.024 & 0.933 & 0.943 & 0.108 & 0.856 \\
 & 800 & -0.003 & -0.001 & 0.001 & 0.004 & 0.006 & 0.012 & 0.942 & 0.953 & 0.107 & 0.806 \\[5pt]
4 & 200 & 0.023 & 0.023 & 0.026 & 0.030 & 0.035 & 0.108 & 0.919 & 0.926 & 0.243 & 0.750 \\
 & 400 & 0.006 & 0.007 & 0.008 & 0.011 & 0.014 & 0.071 & 0.930 & 0.935 & 0.225 & 0.638 \\
 & 800 & 0.002 & 0.004 & 0.006 & 0.009 & 0.013 & 0.049 & 0.955 & 0.963 & 0.223 & 0.386 \\
\hline
\end{tabular}
\end{center}
\end{table}

\section{Real Data Analysis~\label{s:analysis}}

\subsection{Cohort description}


We applied the proposed method to data from EA-IeDEA, which monitored people living with HIV receiving care at $30$ clinics in Kenya. The analytic cohort included 24,372 patients for whom antiretroviral therapy (ART) was initiated. Of these, 15,672 ($64.3\%$) were female, and the median (interquartile range, IQR) age and CD4 cell count at ART initiation were 36.9 (30.4–45.1) years and 178 (75–285) cells/$\mu$L, respectively. During follow-up, 91 patients ($0.4\%$) were passively reported to have died, while 18,439 (75.7$\%$) missed a scheduled clinic visit and were flagged as lost to clinic. Among those lost, 6,608 ($35.8\%$) were successfully traced in the community, and 1,856 ($28.1\%$) were confirmed deceased. The cause of failure remained unobserved for 11,823 ($63.8\%$) patients who were lost to clinic and were not traced. 
The descriptive characteristics of the study cohort are summarized in Table~\ref{t:desc}.

\begin{table}
\caption{Descriptive characteristics the EA-IeDEA study population. (Age: Age at ART initiation; CD4: CD4 cell count at ART initiation; Disclosure: Disclosure of the HIV status)}
\label{t:desc}
\begin{center}
\begin{tabular}{lcccc}
\hline
 & \multicolumn{4}{c}{Cause of failure}  \\ 
 \cmidrule(lr){2-5}
 & \multicolumn{1}{c}{In care (censored)} & \multicolumn{1}{c}{Gap in care} & \multicolumn{1}{c}{Death} & \multicolumn{1}{c}{Missing} \\
 & \multicolumn{1}{c}{($N=5842$)} & \multicolumn{1}{c}{($N=4752$)} & \multicolumn{1}{c}{($N=1955^*$)} & \multicolumn{1}{c}{($N=11823$)} \\ [1ex]
 & \multicolumn{1}{c}{$n$ (\%)} & \multicolumn{1}{c}{$n$ (\%)} & \multicolumn{1}
 {c}{$n$ (\%)} & \multicolumn{1}{c}{$n$ (\%)} \\
\hline
Gender&&&& \\
 \ \ {\it Female}&3770 (64.5)&3044 (64.1)&976 (49.9)&7882 (66.7) \\
 \ \ {\it Male}&2072 (35.5)&1708 (35.9)&979 (50.1)&3941 (33.3) \\
Disclosure&&&& \\
\ \ {\it No}&2244 (38.4)&1719 (36.2)&535 (27.4)&4403 (37.2) \\
\ \ {\it Yes}&3598 (61.6)&3033 (63.8)&1420 (72.6)&7420 (62.8) \\ [3ex]
 & \multicolumn{1}{c}{Median (IQR)} & \multicolumn{1}{c}{Median (IQR)} & \multicolumn{1}
 {c}{Median (IQR)} & \multicolumn{1}{c}{Median (IQR)} \\
 \hline
Age (years)&39.6 (33.2, 47.5)&35.4 (28.8, 42.6)&39.1 (32.3, 48.2)&35.8 (29.3, 43.8) \\
CD4 (cells/$\mu$l)&200 (102, 297)&176 (74, 286)&67 (22, 168)&183 (81, 291) \\
\hline
\multicolumn{5}{l}{\small $^*$ Includes 91 passively reported deaths and 1864 actively ascertained deaths through outreach}
\end{tabular}
\end{center}
\end{table}

\subsection{Model specification}

We fit the cause-specific proportional hazards model described in Section 2, while accounting for dependence between outcomes of patients receiving care in the same clinic (clustering) following prior work on clustered competing risk data with MAR cause of failure~\citep{Zhou23}. We operated under a working-independence assumption along with a version of the influence function to account for clustering in standard errors and $95\%$ confidence bands and intervals. More details on how we accounted for the clustered nature of the data in this setting are presented in Appendix C. Predictors used in the analysis include age and CD4 cell count at the start of ART, plus gender and disclosure of the HIV status.

\subsection{Identification regions}

To compute the identification regions we assumed that the sensitivity parameter $\eta$, which characterizes the association between being untraced and dying (i.e., deviation from the MAR assumption), falls within $[-1, 1]$, corresponding to odds ratios of death for those not traced vs those traced between $e^{-1}=0.37$ and $e^{1}=2.72$. 
Odds ratios over $1$ imply an elevated mortality risk among the lost patients who were not traced in the community compared to those who were traced, while an odds below $1$ implies the opposite. An odds ratio of $1$ is equivalent to the MAR assumption (equal risk of death among patients who were traced and those who were not traced). These odds ratios indicate a range of marginal probabilities of death among the non-successfully traced (i.e., with missing cause of failure) patients, i.e., $\hat{P}(C=2|R=0;\hat{\gamma}_n,\eta)$ with $\eta\in[-1,1]$, from $0.100$ to $0.372$ (marginal probability of death under MAR: $0.210$), which is considered realistic for this application. The estimator of these marginal probabilities is provided in Appendix~D.

Figure~\ref{f:cd4plot} shows the estimated functional cause-specific hazard ratios (HRs) for CD4 cell count as functions of $\eta$, together with the simultaneous $95\%$ confidence bands. Table~\ref{t:analysis1} summarizes identification regions (IRs) and their $95\%$ confidence intervals for all covariates. For the gap-in-care analysis, higher CD4 counts were associated with a slightly increased hazard. The limits of the IR are the minimum and maximum of the estimated functional cause-specific hazard ratio per 100 CD4 cells/$\mu$l (solid blue line) over the range $[-1,1]$ for $\eta$, i.e., $\textrm{IR}=[1.02, 1.06]$, as described in Section~\ref{ss:IR}. The corresponding $95\%$ confidence interval for the IR of the cause-specific hazard ratio of CD4 cell count is $[1.01,1.07]$ in the gap-in-care analysis. Similarly, the CD4 count IR for the mortality cause of failure is $[0.59,0.68]$, with corresponding 95\% confidence interval $[0.56, 0.73]$. For mortality, male sex and lower CD4 counts were strong risk factors (IR = [$1.36$, $1.48$] and [$0.59$, $0.68$]). These associations remained qualitatively stable across the plausible MNAR range, indicating that modest deviations from MAR would not change inference.

\begin{figure}
\begin{center}
\centerline{\includegraphics[width=17cm]{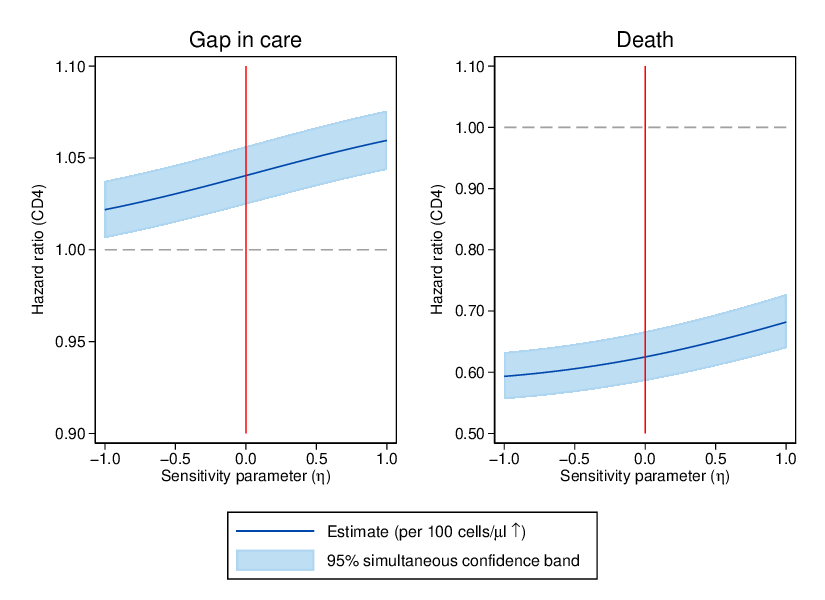}}
\end{center}
\caption{Cause-specific hazard ratios for the effect of CD4 cell count as a function of the sensitivity parameter $\eta$, along with the corresponding 95\% simultaneous confidence bands. The cause-specific hazard ratios at $\eta=0$ correspond to those estimated under the MAR assumption.
\label{f:cd4plot}}
\end{figure}

\begin{table}
\caption{Analysis of the EA-IeDEA data assuming the range $[a,b]=[-1,1]$ for the sensitivity parameter. (IR: Identification region for the cause-specific hazard ratio; CI: Confidence interval for the identification region of the cause-specific hazard ratio; Age: Age at ART initiation; CD4: CD4 cell count at ART initiation; Disclosure: Disclosure of the HIV status)}
\label{t:analysis1}
\begin{center}
\begin{tabular}{lcc}
\hline
 & \multicolumn{1}{c}{IR} &  \multicolumn{1}{c}{95\% CI}  \\ \hline
\underline{Gap in care} && \\
\ \ {\it Gender (male vs. female)} & [0.98, 1.05] & [0.88, 1.17] \\
\ \ {\it Age (per 10 years)} & [0.76, 0.79] & [0.71, 0.84] \\
\ \ {\it CD4 cell count (per 100 cells/$\mu$l)} & [1.02, 1.06] & [1.01, 1.08] \\
\ \ {\it Disclosure (yes vs. no)} & [0.88, 0.90] & [0.79, 1.00] \\[2ex]
\underline{Death} && \\
\ \ {\it Gender (male vs. female)} & [1.36, 1.48] & [1.20, 1.67] \\
\ \ {\it Age (per 10 years)} & [1.01, 1.11] & [0.94, 1.19] \\
\ \ {\it CD4 cell count (per 100 cells/$\mu$l)} & [0.59, 0.68] & [0.56, 0.73] \\
\ \ {\it Disclosure (yes vs. no)} & [1.16, 1.34] & [0.98, 1.57]
 \\
\hline
\end{tabular}
\end{center}
\end{table}


\subsection{Robustness intervals}

To avoid imposing assumptions on the range of $\eta$ and identify the range of MNAR scenarios for which effects remain significant, we computed the robustness intervals. To achieve this, we used the proposed approach (Section~\ref{ss:sensitB}) along with the R function \texttt{uniroot.all} for root finding within the search set $[0,5]$, and set $\epsilon=10^{-8}$. Under this search set, the maximum possible robustness interval was $[-5,5]$, corresponding to odds ratios between $e^{-5}=0.01$ and $e^{5}=148.4$. Table~\ref{t:analysis2} reports robustness intervals for each covariate. The effects of age on gap in care and of CD4 count on mortality remained significant throughout this entire range, suggesting exceptional robustness. Also, the effect of gender on the cause-specific hazard of death remained significant over a wide range of MNAR scenarios, indicating a high degree of robustness against MNAR cause of failure. In contrast, the effects of HIV-status disclosure on gap in care and mortality, as well as the effect of CD4 cell count on gap in care, remained significant within a tighter range of MNAR scenarios, implying a moderate degree of robustness against MNAR. The na\"ive robustness intervals described in Section \ref{ss:sensitB} were considerably more conservative in the gap-in-care analysis, being substantially narrower for CD4 count, and reduced to the empty set for HIV status disclosure. This is indicative of the fact that the na\"ive robustness intervals underestimate the robustness of conclusions regarding statistical hypotheses of interest against MNAR.

\begin{table}
\caption{Results from the MAR analysis of the EA-IeDEA data along with the corresponding proposed and na\"ive robustness intervals. (HR: Cause-specific hazard ratio assuming MAR; CI: Confidence interval assuming MAR; RI: Proposed robustness interval - range of odds ratios of death for those with missing vital status compared to those with known vital status which preserve statistical significance; RI$_{\textrm{na\"ive}}$: Conservative robustness interval derived from the confidence band over the full range of $\eta$ values considered; Age: Age at ART initiation; CD4: CD4 cell count at ART initiation;  Disclosure: Disclosure of the HIV status)}
\label{t:analysis2}
\begin{center}
\begin{tabular}{lccccc}
\hline
& \multicolumn{2}{c}{MAR} & \multicolumn{2}{c}{Sensitivity} \\
\cmidrule(lr){2-3} \cmidrule(lr){4-5}
 & \multicolumn{1}{c}{HR} &  \multicolumn{1}{c}{95\% CI} & \multicolumn{1}{c}{RI$^*$} & \multicolumn{1}{c}{RI$_{\textrm{na\"ive}}^*$}\\
\hline
\underline{Gap in care} &&&& \\
\ \ {\it Gender (male vs. female)} & 1.01 & [0.92, 1.12] & $\emptyset^{**}$ & $\emptyset^{**}$ \\
\ \ {\it Age (per 10 years)} & 0.77 & [0.74, 0.81] & [0.01, 148.41] & [0.01, 148.41]\\
\ \ {\it CD4 cell count (per 100 cells/$\mu$l)} & 1.04 & [1.03, 1.05] &	[0.29, 3.49] & [0.68, 1.47] \\
\ \ {\it Disclosure (yes vs. no)} & 0.89 & [0.82, 0.96] & [0.39, 2.53] & $\emptyset^{**}$ \\[2ex]
\underline{Death} &&&& \\
\ \ {\it Gender (male vs. female)} & 1.44 & [1.28, 1.63] & [0.01, 98.40] & [0.01, 97.61] \\
\ \ {\it Age (per 10 years)} & 1.07 & [0.99, 1.15] & $\emptyset^{**}$ & $\emptyset^{**}$ \\
\ \ {\it CD4 cell count (per 100 cells/$\mu$l)} & 0.63 & [0.59, 0.66] & [0.01, 148.41] & [0.01, 148.41]
 \\
\ \ {\it Disclosure (yes vs. no)} & 1.25 & [1.06, 1.48] & [0.46, 2.17] & [0.46, 2.17]
 \\
\hline
\multicolumn{4}{l}{\small $^*$ Maximum possible robustness interval in this analysis was $[0.01,148.41]$} \\
\multicolumn{4}{l}{\small $^{**}$ Empty set implies a nonsignificant result even under MAR} \\
\end{tabular}
\end{center}
\end{table}

\section{Discussion \label{s:discussion}}

In this research, we proposed a novel sensitivity analysis for the semiparametric proportional cause-specific hazards model with causes of failure being MNAR. We considered a class of MNAR scenarios under a pattern-mixture model specification and provided uniformly consistent estimates of the regression coefficients as functions of a sensitivity parameter (i.e., MNAR scenario). In addition, we proposed a rigorous wild bootstrap procedure for computing simultaneous confidence bands for the functional regression coefficients, capturing the additional uncertainty due to the unknown MNAR scenario within the assumed class. The functional regression coefficient estimator and the simultaneous confidence band are used to estimate the identification region for the true parameter, along with the associated confidence interval, as well as the robustness interval. The latter is defined as the range of MNAR scenarios within which an effect remains statistically significant, and provides a measure of the extent of the robustness of conclusions about statistical hypotheses of interest against MNAR cause of failure. An alternative conservative approach to estimating robustness intervals is to compute the confidence band over the broadest set $[-\eta_{\max},\eta_{\max}]$ and identify the range of $\eta$ values where the band excludes zero. However, this approach tends to yield unnecessarily narrower robustness intervals, as illustrated in our real data analysis, underestimating the robustness of conclusions about statistical hypotheses against MNAR causes of failure. The validity of the proposed methods was established using empirical process theory, while simulation experiments showed a good numerical performance and the robustness of the methods against some degree of mispecification of the assumed pattern-mixture model. To increase flexibility, the assumed model for the probability of cause of failure $C=2$ can incorporate regression splines with a fixed number of internal knots for continuous covariates. 

The proposed approach is related to prior work on sensitivity analysis for general MNAR problems. \citet{Vansteelandt06} provided the definitions and properties of the weak consistency of an identification region estimator, which they call an honestly estimated ignorance region under MNAR. They also introduced the concepts of pointwise, strong, and weak uncertainty region (i.e., confidence interval) for the identification region, for MNAR problems. Our consistency result in Theorem 1 is a stronger version of weak consistency as defined in \citet{Vansteelandt06}, while our proposed confidence interval for the identification region is a strong uncertainty region. \citet{Todem10} and \citet{Cao13} proposed hypothesis testing procedures for parametric and semiparametric models, respectively, under MNAR scenarios. These tests require specifying the range of the sensitivity parameter, which is assumed to be the same across all parameters of interest. In contrast, our proposed robustness intervals do not require this specification, while controlling Type I error in the strong sense, as shown in Section~\ref{ss:sensitB}. In addition, unlike our proposal, these tests evaluate significance over a predetermined sensitivity parameter set and do not provide the range of MNAR scenarios within which statistical significance is preserved. Thus, our proposed robustness intervals provide a more precise way to quantify robustness against MNAR. Applying the tests by \citet{Todem10} and \citet{Cao13} in specific problems, including the one considered in this paper, requires deriving the influence function of the estimator and establishing the necessary entropy conditions \citep[see regularity condition C2 in][]{Cao13} for the problem under consideration. This is not trivial with semiparametric models and typically requires the use of empirical process theory techniques \citep{Kosorok08}. \citet{Zhao19} proposed a percentile bootstrap approach for sensitivity analysis when estimating the mean of an outcome with missing data and the MAR assumption is not warranted. This approach uses inverse probability weighting under a selection model specification to incorporate MNAR scenarios. This sensitivity analysis is not applicable to our problem since we focus on the semiparametric proportional hazards model with competing risks data. Also, we used a pseudolikelihood approach \citep{Bakoyannis20} to deal with missingness instead of inverse probability weighting, since it has been shown to provide substantially more efficient estimates compared to inverse probability weighting in prior simulation studies \citep{Bakoyannis20}.

Beyond missing data, sensitivity analyses have also been developed to assess the robustness of estimates against unmeasured confounding. For instance, \citet{Ding16} and \citet{Vanderweele17} introduced the concept of E-value, a measure quantifying the minimum strength of association that an unmeasured confounder must have with both exposure and outcome to explain away an observed effect. However, the proposed formula for the E-value with survival outcomes relies on the rare-event assumption, in the sense that the survival probability is close to one over the entire follow-up period \citep[see eAppendix 7 in][]{Ding16}, which limits its general use. In contrast, our methodology does not impose assumptions on the rarity of the events under study. Furthermore, there is no formula for computing the E-value in the context of competing risks analysis with MNAR causes of failure, which is addressed in this paper. \citet{Bonvini22} proposed a global robustness summary for average treatment effects against unmeasured confounding. While this approach seeks to identify the most extreme violation of the underlying assumption (i.e., no unmeasured confounding) under which the effect of interest becomes zero, it is not applicable to the problem of competing risks analysis with MNAR causes of failure. Our proposed approach extends the concept of robustness quantification to the competing risks setting, providing a principled method for evaluating the stability of estimates and the robustness of conclusions regarding statistical hypotheses of interest against MNAR causes.


\section*{Acknowledgements}
Research reported in this publication was supported by the National Institute Of Allergy And Infectious Diseases (NIAID), Eunice Kennedy Shriver National Institute Of Child Health \& Human Development (NICHD), National Institute On Drug Abuse (NIDA), National Cancer Institute (NCI), and the National Institute of Mental Health (NIMH), National Institute of Diabetes and Digestive and Kidney Diseases (NIDDK) , Fogarty International Center (FIC), National Heart, Lung, and Blood Institute (NHLBI) , National Institute on Alcohol Abuse and Alchoholism (NIAAA), in accordance with the regulatory requirements of the National Institutes of Health under Award Numbers U01AI069911 East Africa IeDEA Consortium and R21AI145662. The content is solely the responsibility of the authors and does not necessarily represent the official views of the National Institutes of Health.

\bibliographystyle{agsm}  

\bibliography{refs}

\newpage

\section*{Appendices}
\section*{Appendix A: Proofs of Theoretical Results \label{proofs}}
\renewcommand{\thesubsection}{A.\arabic{subsection}}
This Appendix provides the outlines of the proofs of Theorems 1--3 presented in the main text. The proofs rely heavily on empirical process theory techniques \citep{vdVaart96, Kosorok08}. Let $\mathcal{D}$ be the sample space and $D$ a data point in $\mathcal{D}$. Also, let  $P$ be the probability measure in the underlying probability space $(\Omega, \mathcal{F},P)$. For any measurable function $f:\mathcal{D}\mapsto\mathbb{R}$, we use the standard empirical process theory notation 
\[
Pf = \int f\textrm{d}P=E\{f(D)\},
\]
\[
\mathbb{P}_nf = \frac{1}{n}\sum_{i=1}^nf(D_i),
\]
and $\mathbb{G}_nf = \sqrt{n}(\mathbb{P}_n-P)f$. Also, for any probability measure $Q$, define the $L_2$ space 
\[
L_2(Q)=\left\{f:\mathcal{D}\mapsto\mathbb{R}:\|f\|_{Q,2}=\left(\int f^2\textrm{d}Q\right)^{1/2}<\infty\right\}.
\]
To simplify the presentation of the proofs, we introduce some further notation. We use the notation $f\equiv f(\cdot)$ to denote a function $f(x)$ as a whole as opposed to its value for a particular choice of $x$. Also, let $\{C[a,b]\}^p$ denote the space of continuous functions $f:[a,b]\mapsto \mathbb{R}^p$, and define the norm
\[
\|f\|_{[a,b]}\equiv\sup_{\eta\in[a,b]}\|f(\eta)\|, \qquad \mbox{for any} \ f\in\{C[a,b]\}^p.
\]
In addition, let
\[
E(\beta,\eta,t)=\frac{P\{ZY(t)e^{\beta'(\eta)Z}\}}{P\{Y(t)e^{\beta'(\eta)Z}\}},
\]
and 
\[
E_n(\beta,\eta,t)=\frac{\sum_{i=1}^nZ_iY_i(t)e^{\beta'(\eta)Z_i}}{\sum_{i=1}^nY_i(t)e^{\beta'(\eta)Z_i}}.
\] 
Furthermore, let $\phi_{\beta,\eta}(t)=P\{ZY(t)e^{\beta'(\eta)Z}\}$ and $\tilde{\phi}_{\beta,\eta}(t)=P\{Y(t)e^{\beta'(\eta)Z}\}$, so that 
\[
\frac{\phi_{\beta,\eta}(t)}{\tilde{\phi}_{\beta,\eta}(t)}=E(\beta,\eta,t).
\]
Finally, let
\[
H(\beta;\gamma_0,\eta) = \int_0^{\tau}\left(\frac{P\{ZZ'Y(t)e^{\beta'(\eta)Z}\}}{P\{Y(t)e^{\beta'(\eta)Z}\}}-\left[\frac{P\{ZY(t)e^{\beta'(\eta)Z}\}}{P\{Y(t)e^{\beta'(\eta)Z}\}}\right]^{\otimes 2}\right)P\textrm{d}\tilde{N}(t;\gamma_0,\eta).
\]

Our theorems rely on the following regularity conditions:
\begin{itemize}
    \item[C1.] The parameter space for the regression coefficients is
    \[
    \mathcal{B}=\Big\{\beta\in \{C[a,b]\}^p: \|\beta(\eta_1)\|\leq L, \ \ \|\beta(\eta_1)-\beta(\eta_2)\|\leq L|\eta_1 - \eta_2|, \ \ \forall \eta_1,\eta_2\in[a,b]\Big\}
    \]
    for some positive constant $L$. Also, the parameter $\beta^*$ is contained in an open and convex subset of $\mathcal{B}$.
    \item[C2.] The theoretical cumulative baseline hazards
    \[
    \Lambda_j^*(t;\eta)=\int_0^t\frac{dP\tilde{N}_j(s;\gamma_0,\eta)}{P\{Y(s)e^{\beta_j^{*'}(\eta)Z}\}}, \qquad j=1,2, \ \ t\in[0,\tau], \ \ \eta\in[a,b],
    \]
    are continuous functions on $[0,\tau]$. Also, $E\{Y(\tau)|Z\}\geq K_1>0$ almost surely.
    \item[C3.] The covariates $Z$ and auxiliary covariates $A$ are bounded in the sense that there exist a constant $K_2>0$ such that $P(\|Z\|\vee\|A\|\leq K_2)=1$.
    \item[C4.] The inverse $g$ of the link function of the model for $P(C=2|R=1,\Delta=1,W)$ is continuously differentiable on compacts. Also, the parameter space $\Gamma$ of $\gamma$ is a bounded and convex subset of $\mathbb{R}^{p+q+2}$.
    \item[C5. ] The estimator $\hat{\gamma}_n$ is a strongly consistent consistent estimator of $\gamma_0$ and asymptotically linear, that is $\sqrt{n}(\hat{\gamma}_n-\gamma_0)=n^{-1/2}\sum_{i=1}^n\omega_i+o_p(1)$, with the influence functions satisfying $E(\omega)=0$ and $E\|\omega\|^2<\infty$. Also, the empirical versions of the influence functions $\hat{\omega}_i$, $i=1,\ldots,n$, satisfy $n^{-1}\sum_{i=1}^n\|\hat{\omega}_i-\omega_i\|^2=o_p(1)$.
    \item[C6. ] The variance matrix $\textrm{Var}(Z)$ is positive definite and $\min_{j=1,2}E(R\Delta_j|W)>0$ almost surely.
\end{itemize}
Condition C1 implies that the parameter space $\mathcal{B}$ involves continuous functions only. This is guaranteed by the Berge's theorem of the maximum. The conditions of this theorem are satisfied here since the true log partial pseudolikelihood functions,
\[
l_j^*(\beta)=E\left\{\int_0^{\tau}\bigg(\beta^{\prime} Z - \log\left[E\{Y(t)\exp(\beta^{\prime} Z)\}\right]\bigg)\textrm{d}N_{j}(t)\right\}, \qquad j=1,2,
\]
are continuous in both $\beta$ and $\eta$, and the maximizers $\beta_j^*$, $j=1,2$, are unique (this is justified in the proof of Theorem 1 below). The stronger assumption of Lipschitz continuity in condition C1 is reasonable since the objective functions are continuously differentiable with respect to $\beta$ and $\eta$, and not just continuous. Condition C1 also implies that the parameter space is a bounded and convex subset of the Banach space $(\{C[a,b]\}^p,\|\cdot\|_{[a,b]})$. Boundedness and convexity assumptions on the parameter space are standard in the literature of Z-estimation \citep{Kosorok08}. Conditions C2, C3, and C6 (for a given $\eta$) are standard in the survival analysis literature. Also, conditions C4 and C5 are standard in the literature of semiparametric analysis of competing risks data with missing at random cause of failure \citep{Bakoyannis20}. Conditions C4 and C5 are automatically satisfied if the model for $P(C=2|R=1,\Delta=1,W=w;\gamma_0)$ is a correctly specified logistic regression model estimated via maximum likelihood. The goodness of fit of the latter model can be evaluated using the cumulative residual process approach introduced in \citet{Bakoyannis19} and \citet{Bakoyannis20}. Evidence of a lack of fit can be alleviated by imposing a flexible parametric model including regression splines.

Next, we provide a useful lemma that will be utilized in the proofs of our main theorems. 
\begin{lemma} If conditions C1--C3 are satisfied, then the class of functions
\[
\mathcal{F}_1=\left\{\int_0^\tau \frac{\phi_{\beta,\eta}(t)}{\tilde{\phi}_{\beta,\eta}(t)}dN(t):\beta\in\mathcal{B},\eta\in[a,b]\right\},
\]
where $N(t)=\Delta I(X\leq t)$, $t\in[0,\tau]$, is $P$-Donsker.
\end{lemma}

\textbf{Proof.} We have that
\begin{equation}
\int_0^{\tau}\frac{\phi_{\beta,\eta}(t)}{\tilde{\phi}_{\beta,\eta}(t)}dN(t) = \frac{\phi_{\beta,\eta}(X)}{\tilde{\phi}_{\beta,\eta}(X)}\Delta, \label{int_eq}
\end{equation}
for all $\beta\in\mathcal{B}$ and $\eta\in[a,b]$. Next, consider the class
\[
\mathcal{F}_2=\left\{\frac{1}{\tilde{\phi}_{\beta,\eta}}:\beta\in\mathcal{B},\eta\in[a,b]\right\}.
\]
For all $\beta_1,\beta_2\in\mathcal{B}$ and all $\eta_1,\eta_2\in[a,b]$ we have that
\begin{eqnarray*}
    \left|\tilde{\phi}_{\beta_1,\eta_1}(t)-\tilde{\phi}_{\beta_2,\eta_2}(t)\right|&\leq& \left|\tilde{\phi}_{\beta_1,\eta_1}(t)-\tilde{\phi}_{\beta_2,\eta_1}(t)\right|+\left|\tilde{\phi}_{\beta_2,\eta_1}(t)-\tilde{\phi}_{\beta_2,\eta_2}(t)\right| \\
    &\leq& \left|E\left(e^{\beta_1'(\eta_1)Z}-e^{\beta_2'(\eta_1)Z}\right)\right| + \left|E\left(e^{\beta_2'(\eta_1)Z}-e^{\beta_2'(\eta_2)Z}\right)\right| \\
    &\leq& E\left|e^{\beta_1'(\eta_1)Z}-e^{\beta_2'(\eta_1)Z}\right| + E\left|e^{\beta_2'(\eta_1)Z}-e^{\beta_2'(\eta_2)Z}\right| \\
    && \mbox{by Jensen's inequality} \\
    &\leq& K_3E\|Z\|\left\{\|\beta_1(\eta_1)-\beta_2(\eta_1)\|+\|\beta_2(\eta_1)-\beta_2(\eta_2)\|\right\} \\
    && \mbox{by Lipschitz continuity of} \ \  e^x \ \ \mbox{and Cauchy-Schwartz inequality} \\
    &\leq& K_3K_2\left\{\|\beta_1-\beta_2\|_{[a,b]}+\sup_{\beta\in\mathcal{B}}\|\beta(\eta_1)-\beta(\eta_2)\|\right\} \\
    && \mbox{by condition C3} \\
    &\leq& K_3K_2\left\{\|\beta_1-\beta_2\|_{[a,b]}+L|\eta_1-\eta_2|\right\} \\
    && \mbox{by condition C1}, \\
\end{eqnarray*}
for all $t\in[0,\tau]$. This inequality along with conditions C1 and C3 imply that 
for all $\beta_1,\beta_2\in\mathcal{B}$ and all $\eta_1,\eta_2\in[a,b]$
\begin{eqnarray*}
    \left|\frac{1}{\tilde{\phi}_{\beta_1,\eta_1}(t)}-\frac{1}{\tilde{\phi}_{\beta_2,\eta_2}(t)}\right|&\leq& \frac{1}{K_1^2e^{-2pLK_2}}\left|\tilde{\phi}_{\beta_1,\eta_1}(t)-\tilde{\phi}_{\beta_2,\eta_2}(t)\right| \\
    &\leq& K_4\left\{\|\beta_1-\beta_2\|_{[a,b]}+L|\eta_1-\eta_2|\right\}
\end{eqnarray*}
for all $t\in[0,\tau]$, where
\[
K_4 = \frac{K_3K_2}{K_1^2e^{-2pLK_2}}.
\]
The last inequality implies that for any $\epsilon>0$, any $\beta\in\mathcal{B}$, any $\eta\in[a,b]$, and any probability measure $Q$, there exists a $\beta_l\in\mathcal{B}$, $l=1,\ldots,N(\epsilon/(2K_4),\mathcal{B},\|\cdot\|_{[a,b]})$, and an $\eta_m\in[a,b]$, $m=1,\ldots,N(\epsilon/(2LK_4),[a,b],|\cdot|)$, such that
\[
\left\|\frac{1}{\tilde{\phi}_{\beta,\eta}}-\frac{1}{\tilde{\phi}_{\beta_l,\eta_m}}\right\|_{Q,2}\leq \epsilon.
\]
This implies that the class $\mathcal{F}_2$ can be covered by 
\[
N(\epsilon/(2K_4),\mathcal{B},\|\cdot\|_{[a,b]})\times N(\epsilon/(2LK_4),[a,b],|\cdot|)
\]
$L_2(Q)$ $2\epsilon$-balls with centers $\tilde{\phi}_{\beta_l,\eta_m}$, $l=1,\ldots,N(\epsilon/(2K_4),\mathcal{B},\|\cdot\|_{[a,b]})$, $m=1,\ldots,$ \\ $N(\epsilon/(2LK_4),[a,b],|\cdot|)$. Also, note that, by conditions C1 and C3, $|f(t)|\leq F\equiv K_1^{-1}e^{-pLK_2}<\infty$, for all $f\in\mathcal{F}_2$ and $t\in[0,\tau]$. Thus, the uniform covering number of the class $\mathcal{F}_2$ can be bounded as follows
\begin{eqnarray*}
\sup_{Q}N(2\epsilon\|F\|_{Q,2},\mathcal{F}_2,L_2(Q))&\leq& N(\epsilon F/(2K_4),\mathcal{B},\|\cdot\|_{[a,b]})N(\epsilon F/(2LK_4),[a,b],|\cdot|) \\
&\leq& N_{[]}(\epsilon F/(2K_4),\mathcal{B},\|\cdot\|_{[a,b]})N(\epsilon F/(2LK_4),[a,b],|\cdot|) \\
&\leq& N(\epsilon F/(4LK_4),[a,b],|\cdot|)N(\epsilon F/(2LK_4),[a,b],|\cdot|) \\
&\leq& K_5\left(\frac{1}{\epsilon}\right)^2,
\end{eqnarray*}
where the second inequality follows from Lemma 9.18 in \citet{Kosorok08} and the third inequality from condition C1 and Theorem 9.23 in \citet{Kosorok08}, and $K_5$ is a positive constant. The last inequality implies that the uniform entropy integral satisfies
\begin{eqnarray*}
    \int_0^1\sqrt{\log \sup_{Q}N(\epsilon\|F\|_{Q,2},\mathcal{F}_2,L_2(Q))}d\epsilon &\leq & \int_0^1\sqrt{\log(4K_5)+\log\left(\frac{1}{\epsilon^2}\right)}d\epsilon \\
    &\leq& \sqrt{\log(4K_5)} + \int_0^1\sqrt{1+\log\left(\frac{1}{\epsilon^2}\right)}d\epsilon \\
    &\leq& \sqrt{\log(4K_5)} + \int_0^{\infty}u^{1/2}e^{-u/2}du \\
    &=& \sqrt{\log(4K_5)} + \sqrt{2\pi} \\
    &<&\infty .
\end{eqnarray*}
In addition, using arguments similar to those used in page 142 in \citet{Kosorok08}, it can be shown that the classes $\mathcal{F}_{2,\delta}\equiv \{f_1-f_2:f_1,f_2\in\mathcal{F}_2,\|f_1-f_2\|_{P,2}<\delta\}$ and $\mathcal{F}_{2,\infty}^2\equiv \{(f_1-f_2)^2:f_1,f_2\in\mathcal{F}_2\}$ are $P$-measurable for all $\delta > 0$. Therefore, by Theorem 2.5.2 in \citet{vdVaart96}, it follows that that the class $\mathcal{F}_2$ is $P$-Donsker. Using similar arguments to those used for the class $\mathcal{F}_2$, it can be shown that the class
\[
\mathcal{F}_3=\left\{\phi_{\beta,\eta}:\beta\in\mathcal{B},\eta\in[a,b]\right\},
\]
is also $P$-Donsker, with $\sup_{f\in\mathcal{F}_3}\|f\|\leq K_2e^{pLK_2}<\infty$, by conditions C1 and C3. Therefore, the class $\mathcal{F}_1$ is $P$-Donsker by \eqref{int_eq} and Corollary 9.32 in \citet{Kosorok08}, since it is formed by the product of two uniformly bounded $P$-Donsker classes multiplied by a random variable with bounded second moment. This completes the proof of Lemma 1. $\hfill\rule{1ex}{1ex}$

Since the proofs do not depend on the choice of $j$, we will ignore this subindex for notational simplicity and present the proofs for the case where $j=2$.

\subsection{Proof of Theorem 1}
First, we have that
\begin{eqnarray}
    \sup_{\beta\in\mathcal{B}}\left\|\tilde{G}_{n}(\beta;\hat{\gamma}_n,\cdot)-\tilde{G}^*(\beta;\gamma_0,\cdot)\right\|_{[a,b]}&\leq&\sup_{\beta\in\mathcal{B}}\left\|\tilde{G}_{n}(\beta;\hat{\gamma}_n,\cdot)-\tilde{G}_{n}(\beta;\gamma_0,\cdot)\right\|_{[a,b]} \nonumber \\
    && +\sup_{\beta\in\mathcal{B}}\left\|\tilde{G}_{n}(\beta;\gamma_0,\cdot)-\tilde{G}^*(\beta;\gamma_0,\cdot)\right\|_{[a,b]} \nonumber \\
    &\equiv& \sup_{\beta\in\mathcal{B}}\left\|M_{1,n}(\beta,\cdot)\right\|_{[a,b]} + \sup_{\beta\in\mathcal{B}}\left\|M_{2,n}(\beta,\cdot)\right\|_{[a,b]}. \label{consist1}
\end{eqnarray}
For the first term in the right side of \eqref{consist1} we have
\begin{eqnarray*}
    \|M_{1,n}(\beta,\eta)\| &=&\left\|\mathbb{P}_n(1-R)\{g(\hat{\gamma}_n'\tilde{W}+\eta)-g(\gamma_0'\tilde{W}+\eta)\}\int_0^{\tau}\{Z - E_n(\beta,\eta,t)\}dN(t)\right\| \\
    &\leq& \mathbb{P}_n\left|g(\hat{\gamma}_n'\tilde{W}+\eta)-g(\gamma_0'\tilde{W}+\eta)\right|\left\|\int_0^{\tau}\{Z - E_n(\beta,\eta,t)\}dN(t)\right\| \\
    &\leq&\sup_{w\in\mathbb{R}^{p+q}:\|w\|\leq K_2}|\dot{g}(w)|\|\hat{\gamma}_n-\gamma_0\|K_2\mathbb{P}_n\left\|\int_0^{\tau}\{Z - E_n(\beta,\eta,t)\}dN(t)\right\| \\
    &&\mbox{by conditions C3 and C4} \\
    &\leq &\sup_{w\in\mathbb{R}^{p+q}:\|w\|\leq K_2}|\dot{g}(w)|2K_2^2pN(\tau)\|\hat{\gamma}_n-\gamma_0\|.
\end{eqnarray*}
Given that the latter upper bound does not depend on $\beta$ and $\eta$, and since this bound is measurable, it follows that $\sup_{\beta\in\mathcal{B}}\|M_{1,n}(\beta,\cdot)\|_{[a,b]}\overset{as*}\rightarrow 0$, as a consequence of conditions C3--C5. 

Next, for the second term in the right side of \eqref{consist1} we have that
\begin{eqnarray*}
    \|M_{2,n}(\beta,\eta)\| &\leq& \left\|\mathbb{P}_n\int_0^{\tau}\{E_n(\beta,\eta,t)-E(\beta,\eta,t)\}d\tilde{N}(t;\gamma_0,\eta)\right\| \\
    &&+\left\|(\mathbb{P}_n-P)\int_0^{\tau}E(\beta,\eta,t)d\tilde{N}(t;\gamma_0,\eta)\right\|  \\
    &\leq&\sup_{t\in[0,\tau]}\left\|E_n(\beta,\eta,t)-E(\beta,\eta,t)\right\|\mathbb{P}_n\tilde{N}(\tau;\gamma_0,\eta) \\
    &&+\left\|(\mathbb{P}_n-P)\{R\Delta+(1-R)g(\gamma_0'\tilde{W}+\eta)\}\int_0^{\tau}E(\beta,\eta,t)dN(t)\right\|  \\
    &\leq&\sup_{t\in[0,\tau]}\left\|E_n(\beta,\eta,t)-E(\beta,\eta,t)\right\| \\
    &&+\left\|(\mathbb{P}_n-P)\{R\Delta+(1-R)g(\gamma_0'\tilde{W}+\eta)\}\int_0^{\tau}E(\beta,\eta,t)dN(t)\right\|.
    \end{eqnarray*}
Therefore,
\begin{eqnarray}
    \sup_{\beta\in\mathcal{B}}\|M_{2,n}(\beta,\cdot)\|_{[a,b]} &\leq& \sup_{\beta\in\mathcal{B}, t\in[0,\tau]}\left\|E_n(\beta,\cdot,t)-E(\beta,\cdot,t)\right\|_{[a,b]} \nonumber \\
    &&+\sup_{\beta\in\mathcal{B}}\left\|(\mathbb{P}_n-P)\{R\Delta+(1-R)g(\gamma_0'\tilde{W}+\eta)\}\int_0^{\tau}E(\beta,\eta,t)dN(t)\right\|_{[a,b]} \nonumber \\
    &\equiv& \sup_{\beta\in\mathcal{B},  t\in[0,\tau]}\|M_{2,n}'(\beta,\cdot,t)\|_{[a,b]}+\sup_{\beta\in\mathcal{B}}\|M_{2,n}''(\beta,\cdot)\|_{[a,b]}. \label{consist2}
    \end{eqnarray}
For the first term in \eqref{consist2}, note that the fixed classes $\{\beta:\beta\in\mathcal{B}\}$ and $\{\eta:\eta\in[a,b]\}$ are trivially $P$-Donsker, and thus the class $\{e^{\beta'(\eta)Z}:\beta\in\mathcal{B},\eta\in[a,b]\}$ is also $P$-Donsker by Corollary 9.32 (iv) in \citet{Kosorok08}. Thus, the classes $\{ZY(t)e^{\beta'(\eta)Z}:t\in[0,\tau],\beta\in\mathcal{B},\eta\in[a,b]\}$ and $\{Y(t)e^{\beta'(\eta)Z}:t\in[0,\tau],\beta\in\mathcal{B},\eta\in[a,b]\}$ are $P$-Donsker by Lemma 4.1 in \citet{Kosorok08} and the fact that they are formed by the product of two uniformly bounded $P$-Donsker classes. Therefore, the latter two classes are also $P$-Glivenko--Cantelli and thus
\[
\sup_{\beta\in\mathcal{B}, t\in[0,\tau]}\left\|(\mathbb{P}_n-P)ZY(t)e^{\beta'(\cdot)Z}\right\|_{[a,b]}\overset{as*}\rightarrow 0,
\]
and
\[
\sup_{\beta\in\mathcal{B}, t\in[0,\tau]}\left\|(\mathbb{P}_n-P)Y(t)e^{\beta'(\cdot)Z}\right\|_{[a,b]}\overset{as*}\rightarrow 0.
\]
These results, along with the continuous mapping theorem and conditions C1--C3, lead to the conclusion that
\[
\sup_{\beta\in\mathcal{B},  t\in[0,\tau]}\|M_{2,n}'(\beta,\cdot,t)\|_{[a,b]}\overset{as*}\rightarrow 0.
\]
For the second term in \eqref{consist2}, we have that the class of functions 
\[
\left\{\{R\Delta+(1-R)g(\gamma_0'\tilde{W}+\eta)\}\int_0^{\tau}E(\beta,\eta,t)dN(t):\beta\in\mathcal{B},\eta\in[a,b]\right\}
\]
is $P$-Donsker by Corollary 9.32 (iv) in \citet{Kosorok08}, Lemma 1, and the fact that it is formed by the product of two uniformly bounded $P$-Donsker classes. Thus, the latter class is also $P$-Glivenko--Cantelli which implies that
\[
\sup_{\beta\in\mathcal{B}}\|M_{2,n}''(\beta,\cdot)\|_{[a,b]}\overset{as*}\rightarrow 0.
\]
Therefore, by \eqref{consist1} and \eqref{consist2} we have that 
\[
   \sup_{\beta\in\mathcal{B}}\left\|\tilde{G}_{n}(\beta;\hat{\gamma}_n,\cdot)-\tilde{G}^*(\beta;\gamma_0,\cdot)\right\|_{[a,b]}\overset{as*}\rightarrow 0.
\]
Next, to show the uniqueness of the maximizer $\beta^*$, we will first leverage proposition 3.11 in \citet{Peypouquet15}, to show the strict concavity of the true log partial pseudolikelihood function $l^*(\beta)$ for all $\beta\in\mathcal{B}$. The true log partial pseudolikelihood function is twice G{\^a}teaux-differentiable with the second derivative at every $\beta$ in the convex hull of the interior of $\mathcal{B}$ in the direction $(h_1,h_2)$, $h_1,h_2\in\mathcal{B}$ being 
\[
\ddot{l}^*(\beta)(h_1,h_2)=-h_1'H(\beta; \gamma_0, \cdot)h_2. By 
\]
This fact, the result in page 56 in \citet{Kosorok08}, conditions C3 and C6, and proposition 3.11 in \citet{Peypouquet15}, imply the strict concavity of $l^*(\beta)$ over the convex hull of the interior of $\mathcal{B}$. This result, along with the boundedness assumed in conditions C1 and C3, and theorem 2.19 in \citet{Peypouquet15}, lead to the conclusion that the maximizer $\beta^*$ of $l^*(\beta)$ is unique. Therefore, 
\[
\|\hat{\beta}_n-\beta^*\|_{[a,b]}\overset{as*}\rightarrow 0.
\]
This completes the proof of the first statement of Theorem 1. The second statement of Theorem is implied by the first statement and the continuous mapping theorem.

\subsection{Proof of Theorem 2}
By the definition of $\hat{\beta}_n$, we have that 
\begin{eqnarray}
    0&=&\sqrt{n}\tilde{G}_{n}(\hat{\beta}_n(\eta);\hat{\gamma}_n, \eta) \nonumber \\
    &=& \sqrt{n}\left\{\tilde{G}_{n}(\hat{\beta}_n(\eta);\hat{\gamma}_n, \eta)-\tilde{G}_{n}(\hat{\beta}_n(\eta);\gamma_0, \eta)\right\} + \sqrt{n}\tilde{G}_{n}(\hat{\beta}_n(\eta);\gamma_0, \eta) \nonumber \\
    &\equiv& \sqrt{n}M_{1,n}(\hat{\beta}_n,\eta) + \sqrt{n}\tilde{G}_{n}(\hat{\beta}_n(\eta);\gamma_0, \eta), \label{th2_main}
\end{eqnarray}
for all $\eta\in[a,b]$. For the first term in the right size of \eqref{th2_main}, it is straightforward to see that
\begin{eqnarray}
    \sqrt{n}M_{1,n}(\hat{\beta}_n,\eta) &=& \sqrt{n}(\mathbb{P}_n-P)\left[(1-R)\{g(\hat{\gamma}_n'\tilde{W}+\eta)-g(\gamma_0'\tilde{W}+\eta)\}ZN(\tau)\right] \nonumber \\
    && - \sqrt{n}\mathbb{P}_n(1-R)\{g(\hat{\gamma}_n'\tilde{W}+\eta)-g(\gamma_0'\tilde{W}+\eta)\} \nonumber \\
    &&\times\int_0^{\tau}\{E_n(\hat{\beta}_n,\eta,t) - E(\beta^*,\eta,t)\}dN(t) \nonumber \\
    && - \sqrt{n}(\mathbb{P}_n-P)\left[(1-R)\{g(\hat{\gamma}_n'\tilde{W}+\eta)-g(\gamma_0'\tilde{W}+\eta)\}\int_0^{\tau}E(\beta^*,\eta,t)dN(t)\right] \nonumber \\
    && + \sqrt{n}P\left[(1-R)\{g(\hat{\gamma}_n'\tilde{W}+\eta)-g(\gamma_0'\tilde{W}+\eta)\}\int_0^{\tau}\{Z-E(\beta^*,\eta,t)\}dN(t)\right] \nonumber \\
    &\equiv& \epsilon_{1,n}(\eta)-\epsilon_{2,n}(\eta)-\epsilon_{3,n}(\eta) + \sqrt{n}M_{1,n}'(\eta). \label{th2_ineq1}
\end{eqnarray}
First, by conditions C3--C5, the dominated convergence theorem, and a first order Taylor expansion around $\gamma_0$, we have that
\begin{eqnarray*}
    \|\epsilon_{1,n}(\eta)\|&\leq&\left\|(\mathbb{P}_n-P)\left\{(1-R)N(\tau)Z\dot{g}'(\gamma_0'\tilde{W}+\eta)\right\}\sqrt{n}(\hat{\gamma}_n-\gamma_0)\right\|+o_p(1). \\
\end{eqnarray*}
Note that the classes $\{(1-R)N(\tau)Z\}$, $\{\gamma_0'\tilde{W}\}$ and $\{\eta:\eta\in[a,b]\}$ are all trivially $P$-Glivenko--Cantelli. Thus, by condition C4 and Corollary 9.27 (iii) in \citet{Kosorok08}, we have that the class $\{(1-R)N(\tau)Z\dot{g}'(\gamma_0'\tilde{W}+\eta):\eta\in[a,b]\}$ is $P$-Glivenko--Cantelli. Thus, by condition C2 and the central limit theorem, which imply that $\sqrt{n}(\hat{\gamma}_n-\gamma_0)=O_p(1)$, it follows that $\|\epsilon_{1,n}\|_{[a,b]}=o_p(1)$. For the term $\epsilon_{2,n}(\eta)$ note that
\begin{eqnarray}
\|\epsilon_{2,n}\|_{[a,b]}&\leq&\sup_{t\in[0,\tau]}\|E_n(\hat{\beta}_n,\cdot,t) - E(\beta^*,\cdot,t)\|_{[a,b]} \nonumber \\
    &&\times\sqrt{n}\mathbb{P}_n\|(1-R)\{g(\hat{\gamma}_n'\tilde{W}+\cdot)-g(\gamma_0'\tilde{W}+\cdot)\}N(\tau)\|_{[a,b]} \nonumber \\
    &\leq& \sup_{t\in[0,\tau]}\|E_n(\hat{\beta}_n,\cdot,t) - E(\beta^*,\cdot,t)\|_{[a,b]} \nonumber \\
    &&\times K_2\mathbb{P}_n\|\dot{g}'(\gamma_0'\tilde{W}+\cdot)\|_{[a,b]}\|\sqrt{n}(\hat{\gamma}_n-\gamma_0)\|+o_p(1), \label{th2_ineq2}
\end{eqnarray}
by conditions C3--C5, and a first order Taylor expansion around $\gamma_0$. It is not hard to see that
\begin{eqnarray*}
    \|E_n(\hat{\beta}_n,\eta,t) - E(\beta^*,\eta,t)\|&\leq&\|E_n(\hat{\beta}_n,\eta,t) - E_n(\beta^*,\eta,t)\| \\
    && +\left\|\frac{\mathbb{P}_nZY(t)e^{\beta^{*'}(\eta)Z}}{\mathbb{P}_nY(t)e^{\beta^{*'}(\eta)Z}}-\frac{PZY(t)e^{\beta^{*'}(\eta)Z}}{PY(t)e^{\beta^{*'}(\eta)Z}}\right\| \\
    &\leq&\frac{2K_2e^{3pLK_2}}{K_1^2}\mathbb{P}_n|e^{\hat{\beta}_n'(\eta)Z}-e^{\beta^{*'}(\eta)Z}| \\
    && +\left\|\frac{\mathbb{P}_nZY(t)e^{\beta^{*'}(\eta)Z}}{\mathbb{P}_nY(t)e^{\beta^{*'}(\eta)Z}}-\frac{PZY(t)e^{\beta^{*'}(\eta)Z}}{PY(t)e^{\beta^{*'}(\eta)Z}}\right\|
\end{eqnarray*}
by conditions C1--C3. Now, the first term in the right side of the last inequality is independent of $t$ and converges uniformly (in $\eta\in[a,b]$) outer almost surely to 0 by condition C3, the Lipschitz continuity of the exponential function on compacts, the Cauchy--Schwartz inequality, and Theorem 1. For the second term in the right side of the last inequality we have that the classes $\{ZY(t)e^{\beta^{*'}(\eta)Z}:\eta\in[a,b],t\in[0,\tau]\}$ and $\{Y(t)e^{\beta^{*'}(\eta)Z}:\eta\in[a,b],t\in[0,\tau]\}$ are $P$-Donsker by arguments similar to those used in the proof of Theorem 1. Thus, these classes are also $P$-Glivenko--Cantelli and, therefore, it follows from the continuous mapping theorem that
\[
\sup_{t\in[0,\tau]}\left\|\frac{\mathbb{P}_nZY(t)e^{\beta^{*'}(\cdot)Z}}{\mathbb{P}_nY(t)e^{\beta^{*'}(\cdot)Z}}-\frac{PZY(t)e^{\beta^{*'}(\cdot)Z}}{PY(t)e^{\beta^{*'}(\cdot)Z}}\right\|_{[a,b]}\overset{as*}\rightarrow 0.
\]
Consequently, $\sup_{t\in[0,\tau]}\|E_n(\hat{\beta}_n,\cdot,t) - E(\beta^*,\cdot,t)\|_{[a,b]}\overset{as*}\rightarrow 0$. This result along with condition C5, the central limit theorem, and inequality \eqref{th2_ineq2}, imply that $\|\epsilon_{2,n}\|_{[a,b]}=o_p(1)$. Using similar arguments to those used for the term $\epsilon_{1,n}(\eta)$ along with Lemma 1 leads to the conclusion that $\|\epsilon_{3,n}\|_{[a,b]}=o_p(1)$. Finally,
\begin{eqnarray*}
    \sqrt{n}M_{1,n}'(\eta)&=& P\left[(1-R)\int_0^{\tau}\{Z-E(\beta^*,\eta,t)\}dN(t)\dot{g}'(\gamma_0'\tilde{W}+\eta)\right]\sqrt{n}(\hat{\gamma}_n-\gamma_0) + o_p(1) \\
    &=& P\left[(1-R)\int_0^{\tau}\{Z-E(\beta^*,\eta,t)\}dN(t)\dot{g}'(\gamma_0'\tilde{W}+\eta)\right]\frac{1}{\sqrt{n}}\sum_{i=1}^n\omega_i + o_p(1),
\end{eqnarray*}
where the first equality follows from condition C1--C4, the dominated convergence theorem, and a first order Taylor expansion, and the second equality from condition C5.

Next, to analyze the second term in the right side of \eqref{th2_main} we introduce some additional notation. Let
\[
\tilde{\psi}_n(\beta;\eta)=\int_0^{\tau}\left\{Z-E_n(\beta,\eta,t)\right\}d\tilde{N}(t;\gamma_0,\eta),
\]
\[
\tilde{\psi}(\beta;\eta)=\int_0^{\tau}\left\{Z-E(\beta,\eta,t)\right\}d\tilde{N}(t;\gamma_0,\eta),
\]
and
\[
\tilde{M}(t;\beta^*,\eta,\gamma_0)=\tilde{N}(t;\gamma_0,\eta)-\int_0^tY(s)e^{\beta^{*'}(\eta)Z}d\Lambda^*(s;\eta).
\]
Using the latter notation and after some algebra, it can be shown that
\begin{eqnarray}
    \sqrt{n}\tilde{G}_{n}(\hat{\beta}_n(\eta);\gamma_0, \eta) &=& \sqrt{n}\mathbb{P}_n\tilde{\psi}_n(\hat{\beta}_n;\eta) \nonumber \\
    &=& \sqrt{n}\mathbb{P}_n\int_0^{\tau}\left\{Z-E(\beta^*,\eta,t)\right\}d\tilde{M}(t;\beta^*,\eta,\gamma_0) \nonumber \\
    &&+\sqrt{n}\left\{P\tilde{\psi}(\hat{\beta}_n;\eta)-P\tilde{\psi}(\beta^*;\eta)\right\} \nonumber\\
    && + \epsilon_{4,n}(\eta) +\epsilon_{5,n}(\eta) -\epsilon_{6,n}(\eta),\label{th2_psi}
\end{eqnarray}
where
\[
\epsilon_{4,n}(\eta)= \sqrt{n}\mathbb{P}_n\left\{\tilde{\psi}_n(\hat{\beta}_n;\eta) - \tilde{\psi}(\hat{\beta}_n;\eta)-\tilde{\psi}_n(\beta^*;\eta) + \tilde{\psi}(\beta^*;\eta)\right\},
\]
\[
\epsilon_{5,n}(\eta)=\mathbb{G}_n\left\{\tilde{\psi}(\hat{\beta}_n;\eta)-\tilde{\psi}(\beta^*;\eta)\right\},
\]
and
\[
\epsilon_{6,n}=\sqrt{n}\mathbb{P}_n\int_0^{\tau}\left\{E_n(\beta^*,\eta,t)-E(\beta^*,\eta,t)\right\}d\tilde{M}(t;\beta^*,\eta,\gamma_0).
\]
It is straightforward to show that
\begin{eqnarray*}
    \|\epsilon_{4,n}\|_{[a,b]}&\leq& \sup_{t\in[0,\tau]}\left\|\sqrt{n}\left\{E_n(\hat{\beta}_n,\cdot,t)-E(\hat{\beta}_n,\cdot,t)\right\}-\sqrt{n}\left\{E_n(\beta^*,\cdot,t)-E(\beta^*,\cdot,t)\right\}\right\|_{[a,b]} \\
    &\leq&\frac{e^{pLK_2}}{K_1}\bigg[\sup_{t\in[0,\tau]}\left\|\mathbb{G}_nZY(t)\left\{e^{\hat{\beta}_n'(\cdot)Z}-e^{\beta^{*'}(\cdot)Z}\right\}\right\|_{[a,b]} \\
    && +\frac{K_2e^{2pLK_2}}{K_1}\sup_{t\in[0,\tau]}\left\|\mathbb{G}_nY(t)\left\{e^{\hat{\beta}_n'(\cdot)Z}-e^{\beta^{*'}(\cdot)Z}\right\}\right\|_{[a,b]}\bigg],
\end{eqnarray*}
by conditions C1--C3. 
As argued in the proof of Theorem 1, the classes $\{ZY(t)e^{\beta'(\eta)Z}:\beta\in\mathcal{B},\eta\in[a,b],t\in[0,\tau]\}$ and $\{Y(t)e^{\beta'(\eta)Z}:\beta\in\mathcal{B},\eta\in[a,b],t\in[0,\tau]\}$ are $P$-Donsker. In addition,
\begin{eqnarray*}
    \sup_{\eta\in[a,b],t\in[0,\tau]}P\left(Y(t)\left\{e^{\beta'(\eta)Z}-e^{\beta^{*'}(\eta)Z}\right\}\right)^2&\leq&\sup_{\eta\in[a,b]}P\left|e^{\beta'(\eta)Z}-e^{\beta^{*'}(\eta)Z}\right|^2 \\
    &\leq& (K_2K_3)^2\|\beta-\beta^*\|_{[a,b]}^2 \\
    &\rightarrow& 0,
\end{eqnarray*}
as $\|\beta-\beta^*\|_{[a,b]}\rightarrow 0$, where the second inequality follows from the Lipschitz continuity of the exponential function and the Cauchy--Schwartz inequality. Similarly,
\[
 \sup_{\eta\in[a,b],t\in[0,\tau]}P\left\|ZY(t)\left\{e^{\beta'(\eta)Z}-e^{\beta^{*'}(\eta)Z}\right\}\right\|^2\rightarrow 0,
\]
as $\|\beta-\beta^*\|_{[a,b]}\rightarrow 0$. Therefore, by Theorem 1 and arguments similar to those used in Lemma 3.3.5 in \citet{vdVaart96}, it follows that $\|\epsilon_{4,n}\|_{[a,b]}=o_p(1)$. Next, Lemma 1 and condition C4 imply that the class $\{\tilde{\psi}(\beta;\eta):\beta\in\mathcal{B},\eta\in[a,b]\}$ is $P$-Donsker. Furthermore, using similar arguments to those used in the proof of Theorem 1, it can be shown that
\[
\sup_{\eta\in[a,b]}P\left\|\tilde{\psi}(\beta;\eta)-\tilde{\psi}(\beta^*;\eta)\right\|^2\rightarrow 0,
\]
as $\|\beta-\beta^*\|_{[a,b]}\rightarrow 0$. Therefore, by Theorem 1 and arguments similar to those used in Lemma 3.3.5 in \citet{vdVaart96}, it follows that $\|\epsilon_{5,n}\|_{[a,b]}=o_p(1)$.

Next, for $\epsilon_{6,n}(\eta)$, 
condition C1 and Lemmas 4.1 and 15.10 in \citet{Kosorok08} imply that the class $\{\tilde{M}(t;\beta^*,\eta,\gamma_0):t\in[0,\tau],\eta\in[a,b]\}$ is $P$-Donsker. Also, standard arguments imply that $P\{\tilde{M}(t;\beta^*,\eta,\gamma_0)\}=0$. Thus, 
\[
\sqrt{n}\mathbb{P}_n\tilde{M}(\cdot;\beta^*,\cdot,\gamma_0)\leadsto \mathbb{G}_{\tilde{M}} \ \  \mbox{in} \ \ D([0,\tau]\times[a,b]),
\]
where $\mathbb{G}_{\tilde{M}}$ is a tight zero mean Gaussian process and $D([0,\tau]\times[a,b])$ is the space of cadlag functions defined on the set $[0,\tau]\times[a,b]$. In addition, using similar arguments to those used in the proof of Theorem 1 it follows that 
\[
\sup_{t\in[0,\tau]}\left\|E_n(\beta^*,\cdot,t)-E(\beta^*,\cdot,t)\right\|_{[a,b]}\overset{as*}\rightarrow 0.
\]
Thus,
\begin{align*}
    \begin{pmatrix}
        X_n \\
        Y_n \\
    \end{pmatrix}
       &\equiv \begin{pmatrix}
         E_n(\beta^*,\cdot,\cdot)-E(\beta^*,\cdot,\cdot) \\
         \sqrt{n}\mathbb{P}_n\tilde{M}(\cdot;\beta^*,\cdot,\gamma_0) \\
       \end{pmatrix}
    \leadsto \begin{pmatrix}
         0 \\
         \mathbb{G}_{\tilde{M}} \\
       \end{pmatrix}.
\end{align*}
Since the map 
\[
\phi(x,y)(\eta)\equiv\int_0^\tau x(t,\eta)\textrm{d}y(t,\eta),
\]
is continuous at each $(x,y)$ with $x,y\in D([0,\tau]\times[a,b])$ and $\int_0^{\tau}|\textrm{d}x(t,\eta)|<\infty$, it follows from the continuous mapping theorem that
\[
\epsilon_{6,n}\equiv \phi(X_n,Y_n)\leadsto\phi(0,\mathbb{G}_{\tilde{M}})=0.
\]
Note that $\phi(0,\mathbb{G}_{\tilde{M}})(\eta)$ is well defined by integration by parts. Since weak convergence to a constant is equivalent to convergence in probability, we have $\|\epsilon_{6,n}\|_{[a,b]}=o_p(1)$.

Next, it is easy to show that the map $\beta\mapsto \tilde{G}^*(\beta;\gamma_0,\cdot)=P\tilde{\psi}(\beta;\cdot)$ is Fr{\'e}chet differentiable at $\beta^*$ in the direction of $h\in\mathcal{B}$, with derivative $-H(\beta^*;\gamma_0,\cdot)h$, where
\[
H(\beta^*;\gamma_0,\cdot)h = \int_0^{\tau}\left[\frac{PZZ'Y(t)e^{\beta^{*'}Z}}{PY(t)e^{\beta^{*'}Z}}-\left\{\frac{PZY(t)e^{\beta^{*'}Z}}{PY(t)e^{\beta^{*'}Z}}\right\}^{\otimes 2}\right]P\textrm{d}\tilde{N}(t;\gamma_0,\cdot)h, \qquad h\in\mathcal{B},
\]
where $A^{\otimes 2}=AA'$ for any vector $A$. Therefore,
\[
\left\|P\tilde{\psi}(\hat{\beta}_n;\cdot)-P\tilde{\psi}(\beta^*;\cdot) + H(\beta^*;\gamma_0,\cdot)(\hat{\beta}_n-\beta^*)\right\|_{[a,b]} = o_p(\|\hat{\beta}_n-\beta^*\|_{[a,b]}).
\]
Consequently, the second term in the right side of \eqref{th2_psi} we have
\[
\sqrt{n}\left\{P\tilde{\psi}(\hat{\beta}_n;\cdot)-P\tilde{\psi}(\beta^*;\cdot)\right\} = H(\beta^*;\gamma_0,\cdot)\sqrt{n}(\hat{\beta}_n-\beta^*) + o_p(\sqrt{n}\|\hat{\beta}_n-\beta^*\|_{[a,b]}).
\]
By conditions C1 and C6 and the result on page 56 in \citet{Kosorok08}, it follows that $H(\beta^*;\gamma_0,\eta)$ is continuously invertible for all $\eta\in[a,b]$. Thus, using the same arguments as those used on page 311 in \citet{vdVaart96}, it follows that $\sqrt{n}(\hat{\beta}_n-\beta^*)=O_p(1)$, and thus $o_p(\sqrt{n}\|\hat{\beta}_n-\beta^*\|_{[a,b]})=o_p(1)$. Therefore, taking all the pieces together and letting 
\[
\epsilon_n(\eta)=\epsilon_{1,n}(\eta)-\epsilon_{2,n}(\eta)-\epsilon_{3,n}(\eta)+\epsilon_{4,n}(\eta)+\epsilon_{5,n}(\eta)-\epsilon_{6,n}(\eta)+o_p(1),
\]
and 
\begin{eqnarray*}
   \psi_i(\eta)&=&-H^{-1}(\beta^*;\gamma_0,\eta)\Bigg\{\int_0^{\tau}\left\{Z_i-E(\beta^*,\eta,t)\right\}\textrm{d}\tilde{M}_i(t;\beta^*,\eta,\gamma_0) \\
&&+P\left[(1-R)\int_0^{\tau}\{Z-E(\beta^*,\eta,t)\}\textrm{d}N(t)\dot{g}'(\gamma_0'\tilde{W}+\eta)\right]\omega_i\Bigg\},
\end{eqnarray*}
we have by \eqref{th2_main} that
\[
\sqrt{n}\left\{\hat{\beta}_n(\eta) - \beta^*(\eta)\right\}=\frac{1}{\sqrt{n}}\sum_{i=1}^n\psi_i(\eta)+\epsilon_{n}(\eta),
\]
where
\[
\|\epsilon_n\|_{[a,b]}\leq \max_{1\leq l \leq 6}\|\epsilon_{l,n}\|_{[a,b]}+o_p(1)=o_p(1).
\]
Finally, conditions C1, C4, C5, Lemma 1 in Appendix A, and Lemmas 4.1 and 15.10 in \citet{Kosorok08} imply that the class of functions $\{\psi(\eta):\eta\in[a,b]\}$ is $P$-Donsker, which concludes the proof of Theorem 2.

\subsection{Proof of Theorem 3}

The Donsker property is preserved for any linear combination of $P$-Donsker classes of functions. Therefore, for simplicity and without loss of generality, we consider the case of a single covariate, i.e., $p=1$, and $K=1$. Let $BL_1$ denote the space of all Lipschitz functions $f$ with $\|f\|_{\infty}\leq 1$ and Lipschitz constant bounded by 1 \citep{Kosorok08}, and define
\[
\tilde{\mathbb{G}}_n(\eta) = \frac{1}{\sqrt{n}} \sum_{i=1}^n\psi_{i}(\eta)\xi_i,
\]
where $\{\xi_i\}_{i=1}^n$ are i.i.d. standard normal random variables that are independent of the data $\mathcal{D}_n$. Also, let $E_{\xi}$ denote the expectation with respect to $\xi_i$, $i=1,\ldots,n$, conditionally on the observed data $\mathcal{D}_n$. Next, we have
\begin{eqnarray}
    \sup_{h\in BL_1}\left|E_{\xi}\left\{h(\|\hat{\mathbb{G}}_n\|_{[a,b]})\right\}-E\left\{h(\|\mathbb{G}\|_{[a,b]})\right\}\right|&\leq&\sup_{h\in BL_1}\left|E_{\xi}\left\{h(\|\hat{\mathbb{G}}_n\|_{[a,b]})\right\}-E_{\xi}\left\{h(\|\tilde{\mathbb{G}}_n\|_{[a,b]})\right\}\right| \nonumber \\ 
    &&+\sup_{h\in BL_1}\left|E_{\xi}\left\{h(\|\tilde{\mathbb{G}}_n\|_{[a,b]})\right\}-E\left\{h(\|\mathbb{G}\|_{[a,b]})\right\}\right| \nonumber \\
    && \label{WB_ineq}
\end{eqnarray}
For the first term in the right side of inequality \eqref{WB_ineq}, we have by Jensen's inequality that
\begin{equation}
    \sup_{h\in BL_1}\left|E_{\xi}\left\{h(\|\hat{\mathbb{G}}_n\|_{[a,b]})\right\}-E_{\xi}\left\{h(\|\tilde{\mathbb{G}}_n\|_{[a,b]})\right\}\right|\leq\sup_{h\in BL_1}E_{\xi}\left|h(\|\hat{\mathbb{G}}_n\|_{[a,b]})-h(\|\tilde{\mathbb{G}}_n\|_{[a,b]})\right|. \label{WB_ineq_1}
\end{equation}
Note that for any $h\in BL_1$, we have
\begin{eqnarray}
    \left|h(\|\hat{\mathbb{G}}_n\|_{[a,b]})-h(\|\tilde{\mathbb{G}}_n\|_{[a,b]})\right|&\leq&\left|\|\hat{\mathbb{G}}_n\|_{[a,b]}-\|\tilde{\mathbb{G}}_n\|_{[a,b]}\right|\leq\left\|\hat{\mathbb{G}}_n-\tilde{\mathbb{G}}_n\right\|_{[a,b]} \nonumber \\
    &=& \left\|\sqrt{n}\mathbb{P}_n(\hat{\psi} - \psi)\xi\right\|_{[a,b]} \nonumber \\
    &\leq& \left(\left\|\hat{H}_n^{-1}(\hat{\beta}_n;\hat{\gamma}_n,\cdot)-H^{-1}(\beta^*;\gamma_0,\cdot)\right\|_{[a,b]}+\left\|H^{-1}(\beta^*;\gamma_0,\cdot)\right\|_{[a,b]}\right) \nonumber \\
    &&\times\Bigg\{\Bigg\|\sqrt{n}\mathbb{P}_n\Bigg[\int_0^{\tau}\left\{Z-\hat{E}_n(\hat{\beta}_n,\cdot,t)\right\}\textrm{d}\hat{M}(t;\hat{\beta}_n,\cdot,\hat{\gamma}_n) \nonumber \\
    &&-\int_0^{\tau}\left\{Z-E(\beta^*,\cdot,t)\right\}\textrm{d}\tilde{M}(t;\beta^*,\cdot,\gamma_0)\Bigg]\xi\Bigg\|_{[a,b]} \nonumber \\
    &&+\Bigg\|\sqrt{n}\mathbb{P}_n\Bigg(\mathbb{P}_n\left[(1-R)\int_0^{\tau}\{Z-\hat{E}_n(\hat{\beta}_n,\cdot,t)\}\textrm{d}N(t)\dot{g}'(\hat{\gamma}_n'\tilde{W}+\cdot)\right]\hat{\omega} \nonumber \\
    &&-P\left[(1-R)\int_0^{\tau}\{Z-E(\beta^*,\cdot,t)\}\textrm{d}N(t)\dot{g}'(\gamma_0'\tilde{W}+\cdot)\right]\omega\Bigg)\xi\Bigg\|_{[a,b]}\Bigg\} \nonumber \\
    &&+\left\|\hat{H}_n^{-1}(\hat{\beta}_n;\hat{\gamma}_n,\cdot)-H^{-1}(\beta^*;\gamma_0,\cdot)\right\|_{[a,b]}\|H(\beta^*;\gamma_0,\cdot)\|_{[a,b]} \nonumber \\
    &&\times\left\|\sqrt{n}\mathbb{P}_n\psi\xi\right\|_{[a,b]}. \label{WB_ineq_2}
\end{eqnarray}
Using arguments similar to those used in the proof of Theorem 1 along with the continuous mapping theorem lead to the conclusion that
\[
\left\|\hat{H}_n^{-1}(\hat{\beta}_n;\hat{\gamma}_n,\cdot)-H^{-1}(\beta^*;\gamma_0,\cdot)\right\|_{[a,b]}\overset{as*}\rightarrow 0.
\]
In addition, conditions C1-C4 and C6 imply the boundedness of both $\|H^{-1}(\beta^*;\gamma_0,\cdot)\|_{[a,b]}$ and $\|H(\beta^*;\gamma_0,\cdot)\|_{[a,b]}$. Furthermore, using argumes similar to those used in the proof of Theorem 2 as well as in the proof of theorem 3 in \citet{Bakoyannis20} imply that the term inside the curly brackets in the first factor of the right side of \eqref{WB_ineq_2} converges to 0 in probability. Finally, Theorem 2 along with the fact that $\xi\sim N(0,1)$ imply that $\left\|\sqrt{n}\mathbb{P}_n\psi\xi\right\|_{[a,b]}=O_p(1)$. Taking all the pieces together along with inequality \eqref{WB_ineq_2} leads to the conclusion that
\[
\left|h(\|\hat{\mathbb{G}}_n\|_{[a,b]})-h(\|\tilde{\mathbb{G}}_n\|_{[a,b]})\right|\overset{p}\rightarrow 0.
\]
Thus, we have by inequality \eqref{WB_ineq_1}, the definition of $BL_1$, and the dominated convergence theorem, that 
\[
\sup_{h\in BL_1}\left|E_{\xi}\left\{h(\|\hat{\mathbb{G}}_n\|_{[a,b]})\right\}-E_{\xi}\left\{h(\|\tilde{\mathbb{G}}_n\|_{[a,b]})\right\}\right|\overset{p}\rightarrow 0.
\]
For the second term in the right side of \eqref{WB_ineq}, the Donsker property of the class of influence functions by Theorem 2, and theorem 10.4 (conditional multiplier central limit theorem) and proposition 10.7 in \citet{Kosorok08} imply that
\[
\sup_{h\in BL_1}\left|E_{\xi}\left\{h(\|\tilde{\mathbb{G}}_n\|_{[a,b]})\right\}-E\left\{h(\|\mathbb{G}\|_{[a,b]})\right\}\right|\overset{p}\rightarrow 0.
\]
Therefore, it follows that
\[
\sup_{h\in BL_1}\left|E_{\xi}\left\{h(\|\hat{\mathbb{G}}_n\|_{[a,b]})\right\}-E\left\{h(\|\mathbb{G}\|_{[a,b]})\right\}\right|\overset{p}\rightarrow 0,
\]
by \eqref{WB_ineq}. Finally, for any Lipschitz function $\tilde{h}:\mathbb{R}\mapsto [0,1]$ with Lipschitz constant $c_0$ we have
\[
    \left|E_{\xi}\left\{\tilde{h}(\|\hat{\mathbb{G}}_n\|_{[a,b]})\right\}-E\left\{\tilde{h}(\|\mathbb{G}\|_{[a,b]})\right\}\right|\leq c_0\sup_{h\in BL_1}\left|E_{\xi}\left\{h(\|\hat{\mathbb{G}}_n\|_{[a,b]})\right\}-E\left\{h(\|\mathbb{G}\|_{[a,b]})\right\}\right|\overset{p}\rightarrow 0,
\]
and, therefore, by lemma 10.11 in \citet{Kosorok08} it follows that
\[
\sup_{x\in A}\left|P\left\{\|\hat{\mathbb{G}}_{n}\|_{[a,b]}\leq x\bigg|\mathcal{D}_n\right\}-P\left\{\|\mathbb{G}\|_{[a,b]}\leq x\right\}\right|\overset{p}\rightarrow 0, \qquad j=1,2,
\]
for any closed set $A\subset\mathbb{R}\cup\{-\infty,\infty\}$, which concludes the proof.

\section*{Appendix B: Empirical Versions of the Influence Functions \label{infl}}
\renewcommand{\thesubsection}{B.\arabic{subsection}}
To obtain the empirical versions of the influence functions we replace the unknown parameters with their consistent estimators and the sample averages with expectations. That is, the empirical versions of the influence functions are as follows
\begin{eqnarray*}
   \hat{\psi}_{i1}(\eta)&=&-\hat{H}_n^{-1}(\hat{\beta}_{n,j};\hat{\gamma}_n,\eta)\Bigg\{\int_0^{\tau}\left\{Z_i-\hat{E}_n(\hat{\beta}_{n,1},\eta,t)\right\}\textrm{d}\hat{M}_{i1}(t;\hat{\beta}_{n,1},\eta,\hat{\gamma}_n) \\
&&-\mathbb{P}_n\left[(1-R)\int_0^{\tau}\{Z-\hat{E}_n(\hat{\beta}_{n,1},\eta,t)\}\textrm{d}N(t)\dot{g}'(\hat{\gamma}_n'\tilde{W}+\eta)\right]\hat{\omega}_i\Bigg\},
\end{eqnarray*}
and
\begin{eqnarray*}
   \hat{\psi}_{i2}(\eta)&=&-\hat{H}_n^{-1}(\hat{\beta}_{n,j};\hat{\gamma}_n,\eta)\Bigg\{\int_0^{\tau}\left\{Z_i-\hat{E}_n(\hat{\beta}_{n,2},\eta,t)\right\}\textrm{d}\hat{M}_{i2}(t;\hat{\beta}_{n,2},\eta,\hat{\gamma}_n) \\
&&+\mathbb{P}_n\left[(1-R)\int_0^{\tau}\{Z-\hat{E}_n(\hat{\beta}_{n,2},\eta,t)\}\textrm{d}N(t)\dot{g}'(\hat{\gamma}_n'\tilde{W}+\eta)\right]\hat{\omega}_i\Bigg\},
\end{eqnarray*}

where
\[
\hat{H}_{n,j}(\hat{\beta}_{n,j};\hat{\gamma}_n,\eta) = \int_0^{\tau}\left(\frac{\mathbb{P}_n\{ZZ'Y(t)e^{\hat{\beta}_{n,j}'(\eta)Z}\}}{\mathbb{P}_n\{Y(t)e^{\hat{\beta}_{n,j}'(\eta)Z}\}}-\left[\frac{\mathbb{P}_n\{ZY(t)e^{\hat{\beta}_{n,j}'(\eta)Z}\}}{\mathbb{P}_n\{Y(t)e^{\hat{\beta}_{n,j}'(\eta)Z}\}}\right]^{\otimes 2}\right)\mathbb{P}_n\textrm{d}\tilde{N}_j(t;\hat{\gamma}_n,\eta),
\]
\[
\hat{E}_n(\hat{\beta}_n,\eta,t)=\frac{\mathbb{P}_n\{ZY(t)e^{\hat{\beta}_n'(\eta)Z}\}}{\mathbb{P}_n\{Y(t)e^{\hat{\beta}_n'(\eta)Z}\}},
\]
\[
\hat{M}_{ij}(t;\hat{\beta}_{n,j},\eta,\hat{\gamma}_n)=\tilde{N}_{ij}(t;\hat{\gamma}_n,\eta)-\int_0^tY_i(s)e^{\hat{\beta}_{n,j}'(\eta)Z_i}\textrm{d}\hat{\Lambda}_{n,j}(s;\hat{\beta}_{n,j},\eta,\hat{\gamma}_n),
\]
and
\[
\hat{\Lambda}_j(t;\hat{\beta}_{n,j},\eta,\hat{\gamma}_n)=\int_0^t\frac{\mathbb{P}_n\textrm{d}\tilde{N}_{j}(s;\hat{\gamma}_n,\eta)}{\mathbb{P}_nY(s)e^{\hat{\beta}_{n,j}'(\eta)Z}}.
\]

\section*{Appendix C: Marginal Analysis with Clustered Competing Risks Data \label{clust}}
\renewcommand{\thesubsection}{C.\arabic{subsection}}

Consider a clustered-data setting with $n$ clusters (e.g., clinics) and $M_i$ individuals in the $i$th cluster, $i=1,\ldots,n$. Cluster size $M_i$ is considered random and potentially informative \citep{Cong07}, in the sense that it is allowed to be associated with the competing risks data under study. The observed data for the $m$th individual in the $i$th cluster are then $D_{im}=(X_{im},\Delta_{im},\Delta_{im}R_{im}C_{im}, Z_{im}, A_{im}, R_{im})$, $i=1,\ldots,n$, $m=1,\ldots,M_i$. Based on $D_{im}$, let $\Delta_{imj}=I(\Delta_{im}=1,C_{im}=j)$, $j=1,2$, be the indicator that the $m$th individual in the $i$th cluster experienced the $j$th cause of failure. In addition, we define the counting processes $N_{im}(t)=I(X_{im}\leq t, \Delta_{im}=1)$, $t\in[0,\tau]$, and the cause-specific counting process $N_{imj}(t)=I(X_{im}\leq t, \Delta_{imj}=1)=\Delta_{imj}N_{im}(t)$, $j=1,2$. Finally, we define the at-risk process $Y_{im}(t)=I(X_{im}\geq t)$, $t\in[0,\tau]$. In what follows, we assume independence across different clusters but allow for an arbitrary dependence structure within clusters. 

\citet{Zhou23} extended the partial pseudolikelihood approach  for semiparametric analysis of competing risks data under MAR cause of failure \citep{Bakoyannis20} to the clustered-data setting with informative cluster size. To achieve this, they imposed a working-independence assumption and weighted the contribution of each cluster by the inverse of the corresponding cluster size $M_i$. Following \citet{Zhou23} and using the ideas presented in the present paper, estimation of the marginal (or population-averaged) functional regression coefficients can be achieved under a working-independence assumption via the solution of the estimating equation
\[
\tilde{G}_{n,j}(\hat{\beta}_{n,j}(\eta);\hat{\gamma}_n,\eta)=\frac{1}{n}\sum_{i=1}^n\frac{1}{M_i}\sum_{m=1}^{M_i}\int_0^{\tau}\left\{Z_{im} - \hat{E}_n(\hat{\beta}_{n,j},\eta,t)\right\}\textrm{d}\tilde{N}_{imj}(t;\hat{\gamma}_n,\eta)=0, \qquad j=1,2,
\]
where
\[
\hat{E}_n(\beta,\eta,t)=\frac{\sum_{i=1}^nM_i^{-1}\sum_{m=1}^{M_i}Z_{im}Y_{im}(t)e^{\beta'(\eta)Z_{im}}}{\sum_{i=1}^nM_i^{-1}\sum_{m=1}^{M_i}Y_{im}(t)e^{\beta'(\eta)Z_{im}}},
\]
\[
\tilde{N}_{im2}(t;\gamma, \eta) = \{R_{im}\Delta_{im2} + (1 - R_{im})g(\gamma^{\prime}\tilde{W}_{im} + \eta)\}N_{im}(t),
\]
$\tilde{W}_{im}=(1,X_{im},Z_{im}',A_{im}')'$, and 
\[
\tilde{N}_{im1}(t;\gamma, \eta) = [R_{im}\Delta_{im1} + (1 - R_{im})\{1 - g(\gamma^{\prime}\tilde{W}_{im} + \eta)\}]N_{im}(t), \qquad t\in[0,\tau].
\]
The estimate $\hat{\gamma}_n$ is obtained using generalized estimating equations, weighted by the inverse of the cluster size $M_i^{-1}$ to account for a potentially informative cluster size \citep{Cong07}.

The validity of the marginal estimator requires two additional regularity conditions:
\begin{itemize}
    \item[C7.] The cluster size is finite almost surely, that is that there exists a positive integer $m_0$ such that $P(M\le m_0)=1$.
    \item[C8.] The variables $W_{im}$, $\Delta_{im}$, $C_{im}$, and $R_{im}$ are identically distributed conditional on cluster size $M_i$, which implies that $E(W_{im}|M_i)=E(W_{i1}|M_i)$, $E(\Delta_{im}|M_i)=E(\Delta_{i1}|M_i)$, $E(C_{im}|M_i)=E(C_{i1}|M_i)$, and $E(R_{im}|M_i)=E(R_{i1}|M_i)$, for all $i=1,\ldots,n$, $m=1,\ldots,M_i$, and $j=1,2$.
\end{itemize}
The uniform consistency and asymptotic Gaussianity of the estimated functional regression coefficient, as well as the validity of the wild bootstrap, follow from arguments similar to those used in \citet{Zhou23} for competing risks data with MAR cause of failure and the arguments used in the proof of Theorems 1-3 in Appendix A above. The empirical versions of the influence functions for the clustered-data setting are just the cluster-level averages
\[
\hat{\psi}_{ij}(\eta)=\frac{1}{M_i}\sum_{m=1}^{M_i}\hat{\psi}_{imj}(\eta),
\]
where $\hat{\psi}_{imj}(\eta)$ is the influence function of the $m$th individual in the $i$th cluster for the $j$th cause of failure, computed using the formulas provided in Appendix B above. Using this influence function, computation of confidence bands based on wild bootstrap proceeds as described in Section 2.3 of the main text.

\section*{Appendix D: Additional Data Analysis Results \label{sims}}
\renewcommand{\thesubsection}{D.\arabic{subsection}}

To estimate the marginal probabilities of death among the non-successfully traced patients (i.e., with missing cause of failure), $P(C=2|R=0;\gamma_0,\eta)$, in our application we used the following estimator
\[
\frac{\sum_{i=1}^n M_i^{-1}\sum_{m=1}^{M_i}\frac{e^{\hat{\gamma}_n'\tilde{W}_{im}+\eta}}{1+e^{\hat{\gamma}_n'\tilde{W}_{im}+\eta}}(1-R_{im})}{\sum_{i=1}^nM_i^{-1}\sum_{m=1}^{M_i}(1-R_{im})},
\]
for a given sensitivity parameter value $\eta$, where $M_i$ is the number of patients in the $i$th clinic and $\tilde{W}_{im}=(1,X_{im},\textrm{Age}_{im},\textrm{Gender}_{im},\textrm{CD4}_{im},\textrm{Disclosure}_{im})'$, and $X_{im}$ is the event time for the 
$m$th patient in the $i$th clinic. The weight $M_i^{-1}$ in both the numerator and the denominator is used to adjust for the potentially informative cluster size \citep{Zhou23}.

\begin{figure}
\begin{center}
\centerline{\includegraphics[width=17cm]{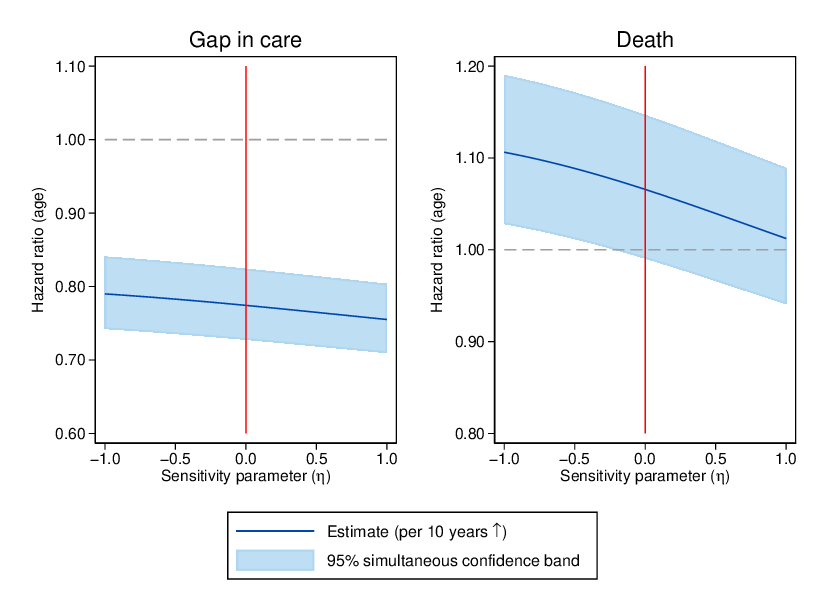}}
\end{center}
\caption{Cause-specific hazard ratios for the effect of age as a function of the sensitivity parameter $\eta$, along with the corresponding 95\% simultaneous confidence bands. The cause-specific hazard ratios at $\eta=0$ correspond to those estimated under the MAR assumption.
\label{f:ageplot}}
\end{figure}

\begin{figure}
\begin{center}
\centerline{\includegraphics[width=17cm]{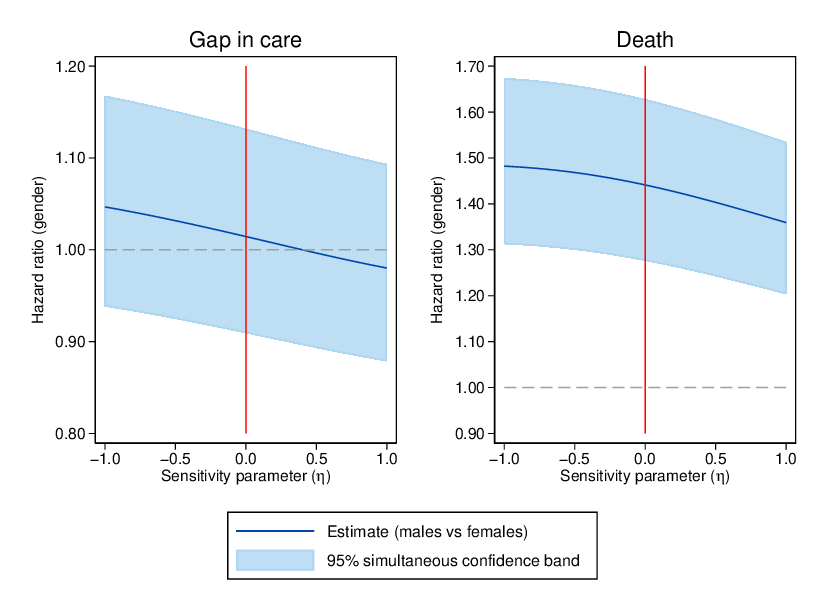}}
\end{center}
\caption{Cause-specific hazard ratios for the effect of gender as a function of the sensitivity parameter $\eta$, along with the corresponding 95\% simultaneous confidence bands. The cause-specific hazard ratios at $\eta=0$ correspond to those estimated under the MAR assumption.
\label{f:maleplot}}
\end{figure}

\begin{figure}
\begin{center}
\centerline{\includegraphics[width=17cm]{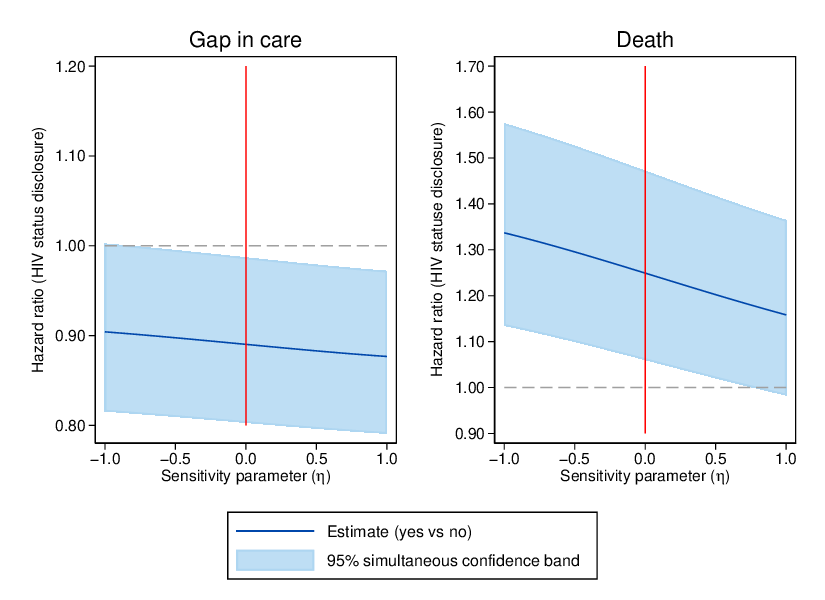}}
\end{center}
\caption{Cause-specific hazard ratios for the effect of HIV status disclosure as a function of the sensitivity parameter $\eta$, along with the corresponding 95\% simultaneous confidence bands. The cause-specific hazard ratios at $\eta=0$ correspond to those estimated under the MAR assumption.
\label{f:statplot}}
\end{figure}

The functional regression coefficients and the corresponding simultaneous 95\% confidence bands for the effects of gender, age, and HIV status disclosure are provided in Figures~\ref{f:ageplot} to \ref{f:statplot}.

\end{document}